\documentclass[12pt]{article}
\renewcommand{\baselinestretch}{1.6}

\usepackage{amsmath}
\usepackage{graphicx}
\usepackage{epstopdf}
\usepackage{hyperref}

\usepackage{times}
\usepackage{bm}
\usepackage{natbib}
\usepackage{amsthm}
\usepackage{amsmath}
\usepackage{amsfonts}

\usepackage[plain,noend]{algorithm2e}

\newtheorem{theorem}{Theorem}
\newtheorem{lemma}{Lemma}

\newcommand{\sd}{\, \begin{picture}(-1,1)(-1,-2)\circle*{2}\end{picture}\ }
\newcommand{\sumi}{\sum_{i=1}^{n}}
\newcommand{\sumii}{\sum_{i=1}^{m}}
\newcommand{\sumj}{\sum_{j=1}^J}
\newcommand{\Kh}[1]{K\left(\frac{#1}{h}\right)}
\newcommand{\cY}{\mathcal{Y}}
\newcommand{\KhX}{\Kh{X_{Pi}-s}}
\newcommand{\fmh}{\frac{1}{mh}}
\newcommand{\cA}{\mathcal{A}}
\newcommand{\cB}{\mathcal{B}}
\newcommand{\cS}{\mathcal{S}}
\newcommand{\cD}{\mathcal{D}}
\newcommand{\cU}{\mathcal{U}}
\newcommand{\cR}{\mathcal{R}}
\newcommand{\cI}{\mathcal{I}}
\newcommand{\cT}{\mathcal{T}}
\newcommand{\pr}{\prime}
\newcommand{\Var}{\mbox{var}}
\newcommand{\Prb}{\mbox{pr}}
\newcommand{\frsm}[2]{\mbox{\small $\frac{#1}{#2}$}}

\def\Var{\mbox{\rm Var}}

\def\Pr{\mbox{\rm Pr}}

\newcommand{\cG}{\mathcal{G}}
\newcommand{\cF}{\mathcal{F}}

\def\bx{{\bar{x}}}

\newcommand{\cC}{\mathcal{C}}
\newcommand{\fC}{\mathfrak{C}}
\newcommand{\fL}{\mathfrak{L}}
\newcommand{\fA}{\mathfrak{A}}
\newcommand{\fB}{\mathfrak{B}}
\newcommand{\cV}{\mathcal{V}}
\newcommand{\cM}{\mathcal{M}}
\newcommand{\cE}{\mathcal{E}}
\newcommand{\cQ}{\mathcal{Q}}
\newcommand{\cP}{\mathcal{P}}

\newcommand{\ocd}{{0,\cD}}

\newcommand{\vs}{\vspace*{6pt}}
\newcommand{\bs}{\bar{s}}
\newcommand{\bt}{\bar{t}}
\newcommand{\real}{\mathbb{R}}

\begin{document}

	\begin{titlepage}
		\vspace*{0.125in}
		\begin{center}
			\LARGE
			\vspace*{1em}
			\bf An improved fully nonparametric estimator \linebreak[4]
			of the marginal survival function \linebreak[4]
			based on case-control clustered data
			\rm
			\normalsize
		\end{center}
		\vspace{1em}
		\begin{center}
			\bf David M. Zucker \\
			\rm Department of Statistics and Data Science, \\
			The Hebrew University of Jerusalem \\
			Mount Scopus, Jerusalem, Israel \\
			\it email\rm: david.zucker@mail.huji.ac.il \\
			\bf and \\
			\bf Malka Gorfine \\
			\rm Department of Statistics and Operations Research \\
			Tel Aviv University, Ramat Gan, Israel \\
			\it email\rm: gorfinem@post.tau.ac.il \\
			\vspace*{1in}
			\today
		\end{center}
	\end{titlepage}
	
	\begin{abstract}
		A case-control family study is a study where individuals with a disease of interest (case probands) and individuals without the disease (control probands)
		are randomly sampled from a well-defined population. Possibly right-censored age at onset and disease status are observed for both probands and their
		relatives. Correlation among the outcomes within a family is induced by factors such as inherited genetic susceptibility, shared environment, and common
		behavior patterns. For this setting, we present a nonparametric estimator of the marginal survival function, based on local linear estimation of
		conditional survival functions. Asymptotic theory for the estimator is provided, and simulation results are presented showing that the method performs
		well. The method is illustrated on data from a prostate cancer study. \\
		
	\end{abstract}
	
	\newpage
	
	\section{Introduction}

Many epidemiological and medical studies focus on disease events that are rare, so that a random sample from the population provides very few observations
where failure has occurred during the monitoring time. In such situations, it is a common practice to consider a case-control strategy. Separate samples
of $n_1$ individuals in whom the event has already occurred (case probands) and $n_0$ individuals in whom the event has not yet occurred (control probands)
are obtained. Age at onset or age at censoring and disease status of each proband and of one or more of his/her relatives are recorded. For example,
in a prostate cancer study, case probands are men diagnosed with prostate cancer, control probands are men without prostate cancer, and each proband
is interviewed to obtain detailed disease history information
of his relatives. The goals of such case-control family studies are to evaluate the effect of genetic and environmental factors on disease risk, and to
estimate the distribution of the age at onset. The analysis of such data is complicated by the case-control selection scheme used to
ascertain the families and by the within-family dependence. It is of interest to estimate the marginal survival function under dependent failure times
of family members by a fully nonparametric estimator which avoids specific assumptions about the form of the distribution or the dependence structure
among failure times with a family.

The estimator that first comes to mind is the Kaplan-Meier survival curve estimator based on the survival data on the relatives. This estimator, however,
is biased because it does not take the case-control sampling
and the within-family dependence into account.
\citet{Gorfine2017} demonstrated the serious bias that can arise with this
naive Kaplan-Meier estimator.

Accordingly, Gorfine et al.\ proposed a nonparametric estimator for this problem, based on a kernel smoothing approach. They presented a simulation study showing that their estimator performs well in terms of bias. The estimator of Gorfine et al. is based on the median of random variables with a complicated dependence structure. Consequently, the asymptotic properties of the estimator were not derived. Also, their use of the median causes some efficiency loss.
In addition, their bandwidth selection procedure was not specifically targeted to the estimand of interest.

In the present paper we develop a new nonparametric estimator which performs well in terms of bias and much better in terms of variance than the estimator of Gorfine et al. In some scenarios the new estimator also outperforms Gorfine et al.\  in terms of bias. We provide asymptotic theory for the estimator. We also present a bandwidth selection procedure that is specifically targeted to the estimand of interest.

\section{Preliminaries}

The setup is as in \citet{Gorfine2017}. We denote the maximum observation time among the probands by $\tau_0$ and the maximum
observation time among the relatives by $\tau$. We write  $n=n_0+n_1$.
We let $J_i$ denote the number of
relatives in family $i$ and we let $J$ denote the maximum number of relatives for a given proband.
We view $J_i$ as a random variable.
We will write some of the formulas as if each proband has exactly $J$ relatives,
with the extra relatives taken to be censored at time 0.
For family $i$, $i=1,\ldots, n$,
let $T_{Pi}$ denote the failure time of the proband and $T_{Rij}, j = 1, \ldots, J_i$
the failure times of the relatives. Let $C_{Pi}$ and $C_{Rij}$ denote the corresponding censoring times, $X_{Pi} =
\min(T_{Pi},C_{Pi})$ and $X_{Rij} = \min(T_{Rij},C_{Pij})$ the corresponding observed times, and
$\delta_{Pi} = I(T_{Pi} \leq C_{Pi})$ and $\delta_{Rij} = I(T_{Rij} \leq C_{Rij})$ the corresponding
event indicators. We assume that the censoring is independent of the survival times.
The sample includes $n_1$ case probands, with each case proband
frequency matched with $a$ control probands, so that the total number of control probands is $n_0=an_1$.
Thus, the data consist of $n_1$ independent and identically distributed matched sets comprising one case
family and $a$ control families, and the observed data on family $i$ consists of
$(X_{Pi},\delta_{Pi},X_{Ri1}, \ldots, X_{RiJ_i},\delta_{Ri1}, \ldots, \delta_{RiJ_i})$.
We seek to estimate the marginal survival function
$$
S(t) = \Prb(T_P>t) = \Prb(T_R>t)
$$
which we are assuming is the same for probands and relatives, a common assumption in case-control
family studies \citep{Shih2002,Chatt2006}.
We further assume that the bivariate
survival function for the proband and a given relative is the same for all relatives. The
marginal survival distribution is assumed to be continuous with density $f(t)$.
In the case where $T_P$ and $T_R$ have different marginal distributions, our procedure
yields a consistent estimate of the marginal distribution of $T_P$.

Denote $S_0(u|t)= \Prb(T_R>u|T_P>t)$ and $S_1(u|t)= \Prb(T_R>u|T_P=t)$. As in Gorfine et al., we have
$\Prb(T_R>u|X_P=t, \delta_P=0) = S_0(u|t)$ and $\Prb(T_R>u|X_P=t, \delta_P=0) = S_1(u|t)$.

We now develop an expression for $S(t)$ in terms of $S_0(u|t)$ and $S_1(u|t)$. Let $\lambda(t)$
and $\Lambda(t)$ denote the hazard and cumulative hazard functions corresponding to $S(t)$, and define
$S_q^*(u|t) = (\partial / \partial t) S_q(u|t)$, $q=0,1$. Also define
$\Lambda_q(u|t) = - \log S_q(u|t)$ and $\Lambda_q^*(u|t) = (\partial / \partial t) \Lambda_q(u|t)$, $q=0,1$.
We can write $S_0^*(u|t) = -S_0(u|t) \Lambda_0^*(u|t)$.
We then have the following:
\begin{align}
& \Prb(T_P>t, T_R>u) = \int_t^\infty \Prb(T_R>u|T_P=x)f(x) dx \nonumber \\
& \Rightarrow \, S(t)S_0(u|t) = \int_t^\infty S_1(u|x)f(x) dx \nonumber \\
& \Rightarrow \, \frac{\partial}{\partial t} [S(t)S_0(u|t)] = -S_1(u|t)f(t) \nonumber \\
& \Rightarrow \, -f(t)S_0(u|t) + S(t)S_0^*(u|t) = -S_1(u|t)f(t) \nonumber \\
& \Rightarrow \,  -\lambda(t)S_0(u|t) + S_0^*(u|t) = -S_1(u|t)\lambda(t) \nonumber \\
& \Rightarrow \,   \lambda(t)(S_0(u|t)-S_1(u|t)) = S_0^*(u|t) = -S_0(u|t)\Lambda_0^*(u|t) \nonumber \\
& \Rightarrow \,   \lambda(t)(S_0(u|t)-S_1(u|t))^2 = -S_0(u|t)(S_0(u|t)-S_1(u|t))\Lambda_0^*(u|t)
\label{pre}
\end{align}
Now define
\begin{equation}
\psi(u,t) = \left[ \int_0^\tau (S_0(v|t)-S_1(v|t))^2 \, dv \right]^{-1} (S_0(u|t)-S_1(u|t))S_0(u|t)
\label{psi}
\end{equation}
Note that the bracketed integral is nonzero provided that for every $t$ there exists a set of $u$ values of positive measure for which
$S_0(u|t) \neq S_1(u|t)$. Integrating both sides of (\ref{pre}) and rearranging gives
\begin{align}
& \lambda(t) =  -\int_0^\tau \psi(u,t) \Lambda_0^*(u|t) \, du \label{small_lam} \\
& \Lambda(t) =  -\int_0^t \int_0^\tau \psi(u,s) \Lambda_0^*(u|s) \, du \, ds
\label{LAM}
\end{align}
We use (\ref{LAM}) to construct our estimator.

\section{Estimation Procedure}

In Gorfine et al., $S_0(u|t)$ and $S_1(u|t)$ were estimated using a generalized version of
the kernel-smoothed Kaplan-Meier estimator proposed in an unpublished 1981 University of
California at Berkeley technical report by R.\ Beran and examined in \citet{Dabr1987},
and the resulting estimators were used to construct an estimator of $S(t)$. Here, in light of (\ref{LAM}), we work not only with $S_0(u|t)$ and $S_1(u|t)$
but also the derivative $\Lambda_0^*(u|t)$. Accordingly, we take a local linear estimation approach. Choose a symmetric kernel function $K$ and
a bandwidth $h$. Let $N_{Rij}(t) = \delta_{Rij}I(T_{Rij} \leq t)$ and $Y_{Rij}(t) = I(T_{Rij}\geq t)$, and write $Q_i = \delta_{Pi}$.
Let $\chi(q_1,q_2) = I(q_1=q_2)$. Define (with $q=0,1$)
\begin{align*}
& \lambda_q(u|t) = \frac{\partial}{\partial u} \Lambda_q(u|t), \quad N_{Ri\sd}(v) = \sumj N_{Rij}(v), \quad
Y_{Ri\sd}(v) = \sumj Y_{Rij}(v) \\
& dM_{Rij}(v) = dN_{Rij}(v) - Y_{Rij}(v) \lambda_{Q_i}(v|X_{Pi}) dv, \quad
M_{Ri\sd}(v) = \sumj M_{Rij}(v) \\
& \cY_q(s,v) =  \frac{1}{n_qh} \sumi \chi(Q_i,q) Y_{Ri\sd}(v) \KhX \\
&  \bar{X}_{Pq}(s,v) = \cY_q(s,v)^{-1} \left[  \frac{1}{n_qh} \sumi \chi(Q_i,q) Y_{Ri\sd}(v) \Kh{X_{Pi}-s}  X_{Pi} \right] \\
& C_q(s,v) = \frac{1}{n_qh} \sumi \chi(Q_i,q) Y_{Ri\sd}(v) \KhX (X_{Pi}-\bar{X}_{Pq}(s,v))^2
\end{align*}
We can write
\begin{align*}
\frac{dN_{Ri\sd}(v)}{Y_{Ri\sd}(v)} & =  \Lambda_{Q_i}(dv|X_{Pi}) + \frac{dM_{Ri\sd}(v)}{Y_{Ri\sd}(v)}
\end{align*}
with $E[dM_{Ri\sd}(v)/Y_{Ri\sd}(v)]=0$.
A first-order Taylor approximation gives the local linear representation
\begin{align*}
\frac{dN_{Ri\sd}(v)}{Y_{Ri\sd}(v)}
& \approx \Lambda_{Q_i}(dv|s) + \Lambda_{Q_i}^*(dv|s)(X_{Pi}-s) + \frac{dM_{Ri\sd}(v)}{Y_{Ri\sd}(v)}
\end{align*}
We now carry out weighted linear least squares with response variable $dN_{Ri\sd}(v)/Y_{Ri\sd}(v)$,
explanatory variable $X_{Pi}-s$, and weights $\chi(Q_i,q)Y_{Ri\sd}(v)K((X_{Pi}-s)/h)$.
This leads to the local linear estimators
\begin{align*}
\widehat{\Lambda}_q^*(dv|s) & = C_q(s,v)^{-1} \left[\frac{1}{n_qh} \sumi \chi(Q_i,q) \KhX (X_{Pi}-\bar{X}_{Pq}(s,v)) dN_{Ri\sd}(v) \right] \\
\widehat{\Lambda}_q(dv|s) & = \frac{(n_qh)^{-1} \sumi \chi(Q_i,q) K((X_{Pi}-s)/h) dN_{Ri\sd}(v)}{\cY_q(s,v)}
- \widehat{\Lambda}_q^*(dv|s) (\bar{X}_{Pq}(s,v)-s)
\end{align*}
that is, for a given $u$,
\begin{align}
\widehat{\Lambda}_q^*(u|s) & = \int_0^u \frac{1}{C_q(s,v)} \left[\frac{1}{n_qh} \sumi \chi(Q_i,q) \KhX (X_{Pi}-\bar{X}_{Pq}(s,v)) dN_{Ri\sd}(v) \right] \\
\widehat{\Lambda}_q(u|s) & = \int_0^u \frac{(n_qh)^{-1} \sum_i \chi(Q_i,q) K((X_{Pi}-t)/h) dN_{Ri\sd}(v)}{\cY_q(s,v)} \nonumber \\
& \hspace*{-36pt} - \int_0^u \frac{\bar{X}_{Pq}(s,v)-s}{C_q(s,v)} \left[\frac{1}{n_qh} \sum_i \chi(Q_i,q) \KhX (X_{Pi}-\bar{X}_{Pq}(s,v)) dN_{Ri\sd}(v) \right]
\end{align}
with the second equation leading to $\widehat{S}_q(u|s) = \exp(-\widehat{\Lambda}_q(u|s))$. We can now substitute
$\widehat{S}_0(u|s)$, $\widehat{S}_0(u|s)$, and $\widehat{\Lambda}_0^*(u|s)$ into (\ref{small_lam}) and (\ref{LAM}) to obtain estimators
$\widehat{\lambda}(t)$ and $\widehat{\Lambda}(t)$ for $\lambda(t)$ and $\Lambda(t)$. We then take $\widehat{S}(t) = \exp(-\widehat{\Lambda}(t))$.
When we want to emphasize the dependence on the bandwidth $h$, we will write $\widehat{S}(t;h)$.

\section{Asymptotic Theory}

We can write
$
\widehat{\Lambda}(t) - \Lambda(t) = \cA(t) + \cA^\pr(t) + \cA^{\pr \pr}(t)
$
with
\begin{align*}
& \cA(t)  = \int_0^t \int_0^\tau \psi(u,s) [\widehat{\Lambda}_0^*(u|s)- \Lambda_0^*(u|s)] \, du \, ds \\
& \cA^\pr(t) = \int_0^t \int_0^\tau [\widehat{\psi}(u,s) - \psi(u,s))] \Lambda_0^*(u|s) \, du \, ds \\
& \cA^{\pr \pr}(t) = \int_0^t \int_0^\tau [\widehat{\psi}(u,s) - \psi(u,s)] [\widehat{\Lambda}_0^*(u|s)- \Lambda_0^*(u|s)] \, du \, ds
\end{align*}
We will provide a detailed asymptotic analysis of $\cA(t)$. Similar arguments can be used to show that $\cA^\pr(t)$ converges in probability
to zero at faster rate than $\cA(t)$, and $\cA^{\pr \pr}(t)$ is negligible in comparison with the other two terms.

In this section and in the appendix, we will write $m=n_0$ and assume that the indices have been arranged so that the control probands
appear first in the list of probands, meaning that a sum over probands 1 to $m$ is a sum over the control probands.

Defining
$$
\Gamma(s,v) = h^{-2} C_0(s,v) = \fmh \sumii  Y_{Ri\sd}(v) \KhX \left(\frac{X_{Pi}-\bar{X}_{P0}(s,v)}{h}\right)^2
$$
we have
\begin{align*}
\cA(t) = - \,\frac{1}{mh^3} \sumii \int_0^t \int_0^\tau \int_0^u & \psi(u,s) \Gamma(s,v)^{-1}
\KhX (X_{Pi}-\bar{X}_{P0}(s,v)) \\
& (dN_{Ri\sd}(v) - Y_{Ri\sd}(v)(X_{Pi}-\bar{X}_{P0}(s,v))\lambda_0^*(v|s) dv) \, du \, ds
\end{align*}
Now, the process $M_{Rij}$ is a martingale with respect to the filtration
$\mathcal{F}_{ijv} = \sigma(X_{Pi},$ $\{N_{Rij}(d),$ $Y_{Rij}(d)$, $d \in [0,v]\})$.
Accordingly, for any process $\mathcal{P}(v)$
that is predictable with respect to $\mathcal{F}_{ijv}$, the process
$$
\int_0^u \mathcal{P}(v) dM_{Rij}(v)
$$
is a mean-zero martingale. It follows, even though the process $M_{Ri\sd}$ does not have any martingale properties,
that for any function $P(v,x)$ we have
\begin{equation}
E \left[ \int_0^u P(v,X_{Pi}) dM_{Ri\sd}(v) \right] = 0
\label{mz}
\end{equation}

We now write $\cA(t) = -(\cA_1(t) + \cA_2(t))$, where
\begin{align*}
& \cA_1(t) = \frac{1}{mh^3} \sumii \int_0^t \int_0^\tau \int_0^u \psi(u,s) \Gamma(s,v)^{-1} Y_{Ri\sd}(v) \KhX (X_{Pi}-\bar{X}_{P0}(s,v))  \\
& \hspace*{100pt} (\lambda_0(v|X_{Pi}) - (X_{Pi}-\bar{X}_{P0}(s,v))\lambda_0^*(v|s)) dv \, \, du \, ds \\
& \cA_2(t) = \frac{1}{mh^3} \sumii \int_0^t \int_0^\tau \int_0^u \psi(u,s)\Gamma(s,v)^{-1} \KhX (X_{Pi}-\bar{X}_{P0}(s,v))
dM_{Ri\sd}(v) \, du \, ds
\end{align*}

In the appendix it is shown, under the conditions listed in the appended supplementary document that $(mh)^{1/2} \cA_1(t) \rightarrow 0$ in probability
and that
$$
\cA_2(t) = \frac{1}{m} \sumii \zeta_i(t) + o_p((mh)^{-1/2})
$$
with
$$
\zeta_i(t) = \frac{1}{h} \int_0^t \int_0^\tau \int_0^u \psi(u,s)\gamma(s,v)^{-1}
\left(\frac{X_{Pi}-s}{h}\right)
\left[h^{-1}\KhX\right] dM_{Ri\sd}(v) \, du \, ds
$$
where $\gamma$ is the limiting value of $\Gamma$. Define $\sigma_{\zeta}^2(t) = h \, \Var(\zeta_i(t))$.
We then obtain the following theorem.

\begin{theorem}
For each $t \in [0,\tau_0]$, $\sigma_{\zeta}^2(t)$ converges to a limit $\kappa(t)$ and $(mh)^{1/2} \, (\widehat{\Lambda}(t) - \Lambda(t))$
converges in distribution to $N(0,\kappa(t))$.
\end{theorem}

The proof of this theorem appears in the appendix. Details are given in the appended supplementary document.

\section{Practical Implementation Details}

In preliminary work, we found that the performance of the estimator of $\widehat{S}(t)$ can be improved dramatically by introducing a time transformation
that makes the proband observation times approximately uniformly distributed. Along the lines of \citet{Doksum2017}, we propose transforming
according to an estimate of the distribution function of the proband observation times, which leads to a modified form of nearest neighbor regression. In the appended supplementary document, we show
that the consistency and asymptotic normality is maintained under the time transformation
if a smooth estimate of the distribution function is used. We believe that this result holds as well when the empirical distribution function is used.
In our numerical work, we used the empirical distribution function.

In the context of family survival data, it is usually reasonable to assume that $\Prb(T_P>t, T_R>u) \geq \Prb(T_P>t) \Prb(T_R>u) = S(t)S(u)$ for
all $t$ and $u$, i.e., $S_0(u|t) \geq S(u)$ for all $t$ and $u$.
This condition implies that
\begin{equation}
\Prb(T_R>u|T_P \leq C_P) \leq S(u) \leq \Prb(T_R>u|T_P > C_P)
\label{bnds}
\end{equation}
If we let $\widehat{S}_{KM,case}(t)$ and $\widehat{S}_{KM,control}(t)$ denote the Kaplan-Meier survival curve estimator based
on the case relatives' survival data and the control relatives' survival data, respectively, the foregoing inequalities
motivate modifying the estimator to the estimator $\tilde{S}(t)$ resulting from replacing $\widehat{S}(t)$ with $\widehat{S}_{KM,case}(t)$ if
$\widehat{S}(t) \leq \widehat{S}_{KM,case}(t)$ and by $\widehat{S}_{KM,control}(t)$ if $\widehat{S}(t) \geq \widehat{S}_{KM,control}(t)$.
We implemented this modification
in our numerical work. The modification comes into play mainly when the event rate is extremely low or extremely high.
Given the consistency of $\widehat{S}(t)$,
if the inequalities in (\ref{bnds}) are strict, then
for large sample sizes the modification no longer comes into play.
The inequalities in (\ref{bnds}) are strict if the following mild condition holds:
for each $u$ there exists a set $\cT(u)$ such that $P(C_P \in \cT(u)) > 0$ and
$$
\inf_{t \in \cT(u)} P(T_P > t, T_R > u)  -  P(T_P > t)P(T_R > u) > 0
$$

For bandwidth selection, we propose a bootstrap procedure. Let $\widehat{S}^C(u)$ denote the Kaplan-Meier estimate of the survival function
of the time to censoring among the relatives (which is the same for case relatives and control relatives). In each bootstrap replication $b=1, \ldots, B$,
for each family $i$ we generate $J_i$ event times for proband $i$'s relatives according to the survival function $\widehat{S}_{Q_i}(u|T_{Pi})$
and $J_i$ censoring times for relatives according to the survival function $\widehat{S}^C(u)$. We then run our estimation procedure for a given $h$ on the resulting data,
obtaining the estimate $\widehat{S}(u;h,b)$. Denote
\begin{align*}
\bar{S}(t;h) & = \frac{1}{B} \sum_{b=1}^B \widehat{S}(t;h,b), \quad
V(t;h) = \frac{1}{B-1} \sum_{b=1}^B (\widehat{S}(t;h,b) - \bar{S}(t;h))^2 \\
\mbox{\textsc{mse}}_{est}(t,h) & = (\bar{S}(t;h) - \widehat{S}(t;h))^2 + V(t;h), \quad
\mbox{\textsc{imse}}_{est}(h) = \int_0^\tau \mbox{\textsc{mse}}_{est}(t,h) dt
\end{align*}
We evaluate $\mbox{\textsc{imse}}_{est}(h)$ over a grid of $h$ values in the range $(0,1]$ and choose the $h$ values with the minimum
$\mbox{\textsc{imse}}_{est}(h)$.

To construct confidence intervals in finite samples with bandwidth selection, we use the percentile bootstrap method.

\section{Simulation Study}

We carried out a simulation study to evaluate the finite sample properties of the proposed estimator. Data were
generated under frailty models in which the within-family dependence is expressed in terms of a shared frailty
variate $W_i$, conditional on which the failure times of the family members are independent with hazard function
$\lambda(t|W_i) = W_i \lambda_0(t)$. We manipulated five design factors, as follows: (1) frailty distribution: gamma or
positive stable, (2) cumulative end-of-study event rate: high (60\%) or low (15\%), (3) number of case probands:
500 or 1000 (with 1:1 matching of control probands to case probands), (4) number of relatives per family: 1 or 4,
and (5) strength of within-family dependence: low (Kendall tau of 1/3 between the failure of times of two members
of the same family) and moderate (Kendall tau of 1/2). We took $\lambda_0(t) = \nu (\mu t)^{p-1}$ with $p=4.6$,
$\mu = 0.01$, and $\nu$ chosen so as to obtain the desired cumulative end-of-study event rate. The end of study age was taken to be 110 years.
The overall censoring rate, including both interim and end-of-study censoring, was about 60\% in the high event
rate case and 90\% in the low event rate case.
The number of case probands was taken to be 500, with 1:1 matching of control probands to case probands.
The data generation was carried out in the same manner as in Gorfine et al.\ We carried out 1024 simulation
replications for each of the 32 combinations of the design. For each replication, we carried out
30 inner replications for the bootstrap bandwidth selection procedure and 100 outer replications for the
percentile bootstrap confidence interval procedure. The initial bandwidth was 0.5
and the bandwidth search was done in two stages. In the first stage, we searched
over $[0.1,1]$ in steps of 0.1 and identified the $h$ value $h_1$ with the lowest $\mbox{\textsc{imse}}_{est}(h)$. In the second stage, we searched over
$h_1-0.05, h_1$, and $\min(h_1+0.05,1)$ and chose the $h$ value with
the lowest $\mbox{\textsc{imse}}_{est}(h)$ to be the final $h$ value.
The kernel used was the triweight kernel $K(u) = (35/32) I(|u|\leq 1)(1-u^2)^3$.

\begin{figure}
\includegraphics[width=\linewidth]{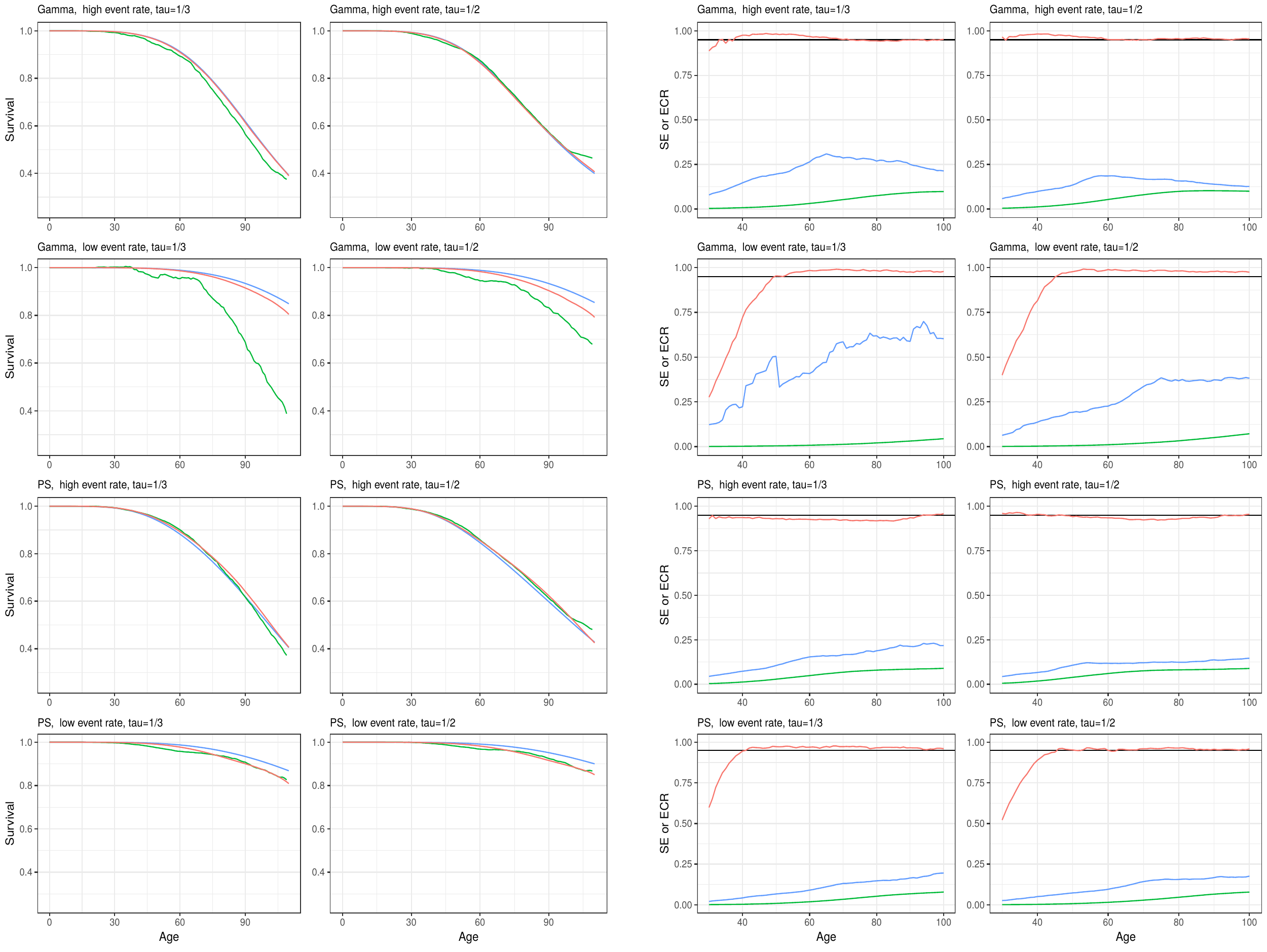}
\caption{Simulation results, one relative for each proband: Left two columns include the true survival curve (blue); Gorfine et al. estimator (green); and the proposed estimators (red). Right two columns present the empirical standard errors of Gorfine et al. (blue) and the proposed estimator (green); and point-wise precentile-bootstrap $95\%$ confidence interval  coverage rates of the proposed estimator. The black horizontal line at 0.95 serves as a reference.}
\label{fig1}
\end{figure}

\begin{figure}
\includegraphics[width=\linewidth]{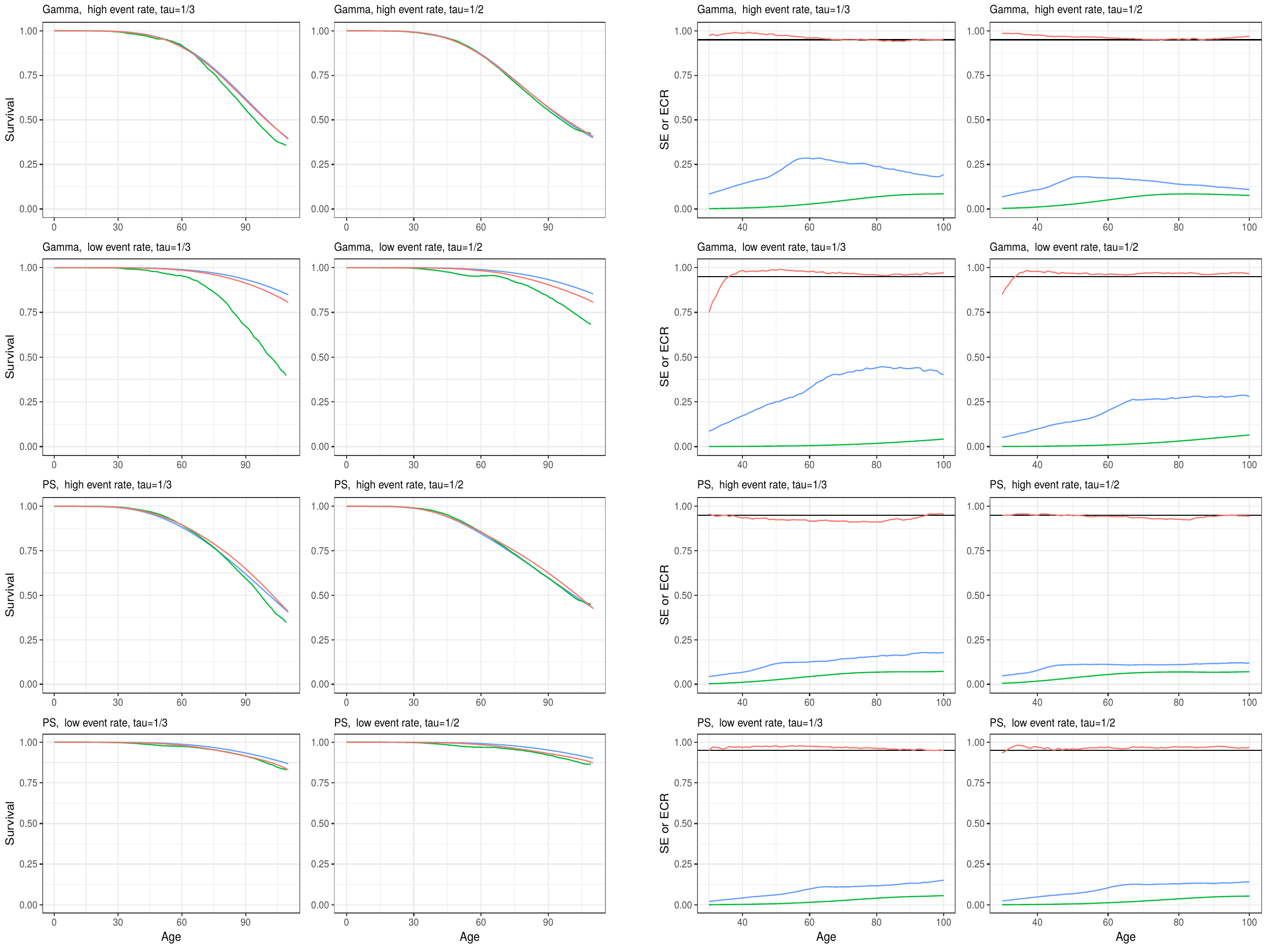}
\caption{Simulation results, four relatives for each proband: Left two columns include the true survival curve (blue); Gorfine et al. estimator (green); and the proposed estimators (red). Right two columns present the empirical standard errors of Gorfine et al. (blue) and the proposed estimator (green); and point-wise precentile-bootstrap $95\%$ confidence interval  coverage rates of the proposed estimator. The black horizontal line at 0.95 serves as a reference.}
\label{fig2}
\end{figure}

The results for 500 case probands are summarized in Fig.~\ref{fig1} and~\ref{fig2}. The left two columns of each figure show the true survival curve, along with Gorfine et al.'s estimator and the new estimator. The finite-sample bias of the new estimator tends to be smaller, and in some settings, such as the gamma frailty model with very low event rates, its finite-sample bias is dramatically smaller. The right two columns of each figure summarize the point-wise $95\%$  coverage rates of the percentile-bootstrap  confidence interval of the proposed estimator, along with the standard errors of the Gorfine et al.\ estimator and the proposed estimators. Clearly, the proposed estimator substantially outperforms the old estimator in terms of efficiency. In general, the coverage rates are reasonably close to $95\%$, except at very early ages with small number of observed events. Similar results were obtained with 1000 case probands.

\renewcommand{\baselinestretch}{1}

\renewcommand{\baselinestretch}{1.6}

\section{Example}

In this section we illustrate our method by re-analysing the data presented in \citet{Gorfine2017},
population-based case-control family study of early onset prostate
cancer \citep{Stan1999}. Briefly, case participants were identified from the Seattle-Puget Sound Surveillance,
Epidemiology, and End Results (SEER) cancer registry. Cases were those
with age at diagnosis between
40 and 64 years. Controls were identified by use of random-digit dialing
and they were frequency matched to case participants by age. The information
collected on the relatives is the age at diagnosis for prostate cancer
if the relative had prostate cancer or age at the last observation
if the relative did not have prostate cancer.  Here we use the information about age at onset
or age at censoring and disease status that was observed for the probands and their fathers, brothers, and
uncles. The following analysis is based on 730 prostate-cancer case probands, 693 control probands, and
a total of 7316 relatives. Out of the 3793 case-probands’ relatives, 211 had prostate cancer, and out of
the 3523 control-probands’ relatives, 102 had prostate cancer. The age range of the relatives with prostate cancer was 40--93. Given that frequency matching was used
rather than exact age matching, and that the number of relatives per proband varied
across the probands, we carried out the time transformation based on the
empirical distribution of the proband observation times across all 7316 relatives
in the data set. For bandwidth selection we used the same two-stage procedure
as in the simulations.

\begin{figure}
\includegraphics[height=10cm,width=\linewidth]{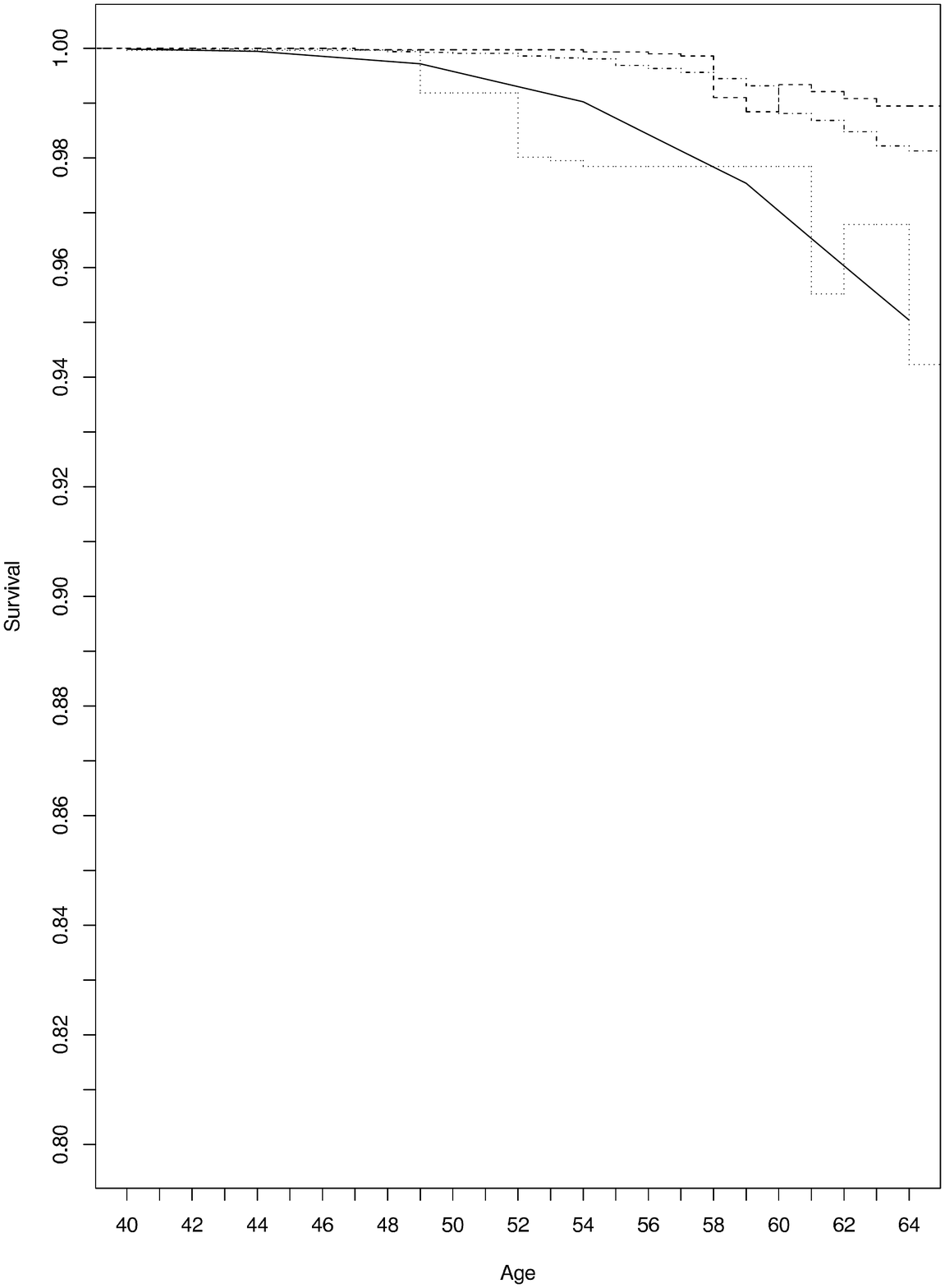}
\caption{Prostate cancer case-control family data: the naive Kaplan-Meier estimator based on the relatives' data (dot-dashed line), Gorfine et al.'s estimator (dotted line), the SEER survival curve (solid line), and the proposed estimator (dashed line).}
\label{fig3}
\end{figure}

Figure \ref{fig3} and Table \ref{tablelabel} present the estimates of prostate-cancer marginal survival function using the naive Kaplan-Meier estimator based on the relatives' data, Gorfine et al.'s estimator with nearest-neighbor smoothing and the median operator, the SEER survival curve based on the SEER Cancer Statistics Review 1975--2012, and the proposed estimator. In this dataset, Gorfine et al.'s estimator is closer to the SEER survival curve, but with very large point-wise standard errors compared to the proposed estimator. The standard errors reported in Table 1 are much larger than those reported in Gorfine et al.\ due to an error in the bootstrap code applied back then.

\renewcommand{\baselinestretch}{1}

\begin{table}
\def~{\hphantom{0}}
\caption{Prostate cancer case-control family data}
\begin{tabular}{lcccc}
 t & The proposed Estimator & Gorfine et al. & Naive KM & SEER \\[5pt]
50 & 0.9997 (0.0007) & 0.9918 (0.0311) & 0.9991 (0.0003) & 0.9958 \\
52 & 0.9997 (0.0010) & 0.9801 (0.0340) & 0.9986 (0.0005) & 0.9930 \\
54 & 0.9993 (0.0012) & 0.9784 (0.0413) & 0.9981 (0.0006) & 0.9902 \\
56 & 0.9990 (0.0023) & 0.9784 (0.0451) & 0.9963 (0.0008) & 0.9843 \\
58 & 0.9910 (0.0037) & 0.9784 (0.0451) & 0.9945  (0.0010) & 0.9783 \\
60 & 0.9934 (0.0054) & 0.9784 (0.0461) & 0.9881 (0.0015) & 0.9703 \\
62 & 0.9908 (0.0063) & 0.9678 (0.0501) & 0.9848 (0.0017) & 0.9603 \\
64 & 0.9895 (0.0085) & 0.9423 (0.0577) & 0.9813 (0.0019) & 0.9504 \\
\end{tabular}
\label{tablelabel}
\end{table}

\renewcommand{\baselinestretch}{1.6}

\section*{Acknowledgements}
The authors would like to thank Dr.\ Stanford for her generosity in sharing the prostate cancer dataset that was used for illustrating the method.
Malka Gorfine gratefully acknowledges support from the U.S.-Israel Binational Science Foundation in carrying out this work.

\section*{Supplementary material}
\label{SM}
We append a document with details of the proof of the asymptotic properties
of the estimator.
R code used to carry out the simulations and R code for applying the method to a data set may be found at
\url{https://github.com/david-zucker/marginal-survival/}.
\newpage

\appendix

\section*{Appendix: Asymptotic Theory}


\noindent
A. \it Preliminaries \rm

We first present some definitions. Let $g(x)$ denote the density of $X_{Pi}$.
Define $y(s,v) = E[Y_{Ri\sd}(v)|X_{Pi}=s, \delta_{Pi}=0]$ and $\varphi(s,v) = g(s)y(s,v)$.
In addition, define
\begin{align*}
& \cI =  [h,\tau_0-h], \quad \cU = [0,h) \cup (\tau_0-h,\tau_0], \quad
Z_k(r) = r^k K(r), \quad \mu_k(\omega) = \int_{-1}^\omega Z_k(r)  dr \\
& A_k(s,v,h) = \fmh \sumii Y_{Ri\sd}(v) Z_k\left(\frac{X_{Pi}-s}{h}\right)  \\
& \eta_k(s,h) = \int_{-s/h}^{(\tau_0-s)/h} Z_k(r) dr
=
\left\{
\begin{array}{ll}
(-1)^k \mu_k(s/h) & s \in [0,h] \\
\mu_k(1) &  s \in \cI \\
\mu_k((\tau_0-s)/h) & s \in [\tau_0-h,\tau_0]
\end{array} \right. \\
& a_k(s,v,h) = \eta_k(s,h) \varphi(s,v)
\end{align*}
Note that, by symmetry of $K$, $\mu_k(1)=0$ for all odd $k$.

We now present two lemmas, whose proofs appear in the online Supplemental Materials.
The notations $O$ and $o$, and similarly $O_p$ and $o_p$, should be understood as being uniform in the relevant
arguments.
\begin{lemma}
For $k$ even we have
$$
\sup_{s,v} |A_k(s,v,h) - a_k(s,v,h)|
= \left\{
\begin{array}{ll}
O_p(m^{-(1-\nu)/2} \, (\log m)^{1/2}) & s \in \cI \\
O_p(m^{-\nu}) & s \in \cU
\end{array}
\right.
$$
and for $k$ odd we have
$$
\sup_{s,v} |A_k(s,v,h)| = O_p(h) \mbox{ for } s \in \cI, \quad
\sup_{s,v} |A_k(s,v,h) - a_k(s,v,h)| =  O_p(h) \mbox{ for } s \in \cU
$$
\end{lemma}

\begin{lemma}
For $s \in \cI$ we have
\begin{align*}
\bar{X}_{P0}(s,v) - s = O_p(h^2), \quad
\Gamma(s,v) = a_2(s,v,h) + O_p(m^{-(1-\nu)/2}  \, (\log m)^{1/2})
\end{align*}
and for $s \in \cU$ we have
\begin{align*}
\bar{X}_{P0}(s,v) - s = \left(\frac{a_1(s,v,h)}{a_0(s,v,h)}\right)h + O_p(h^2), \quad
\Gamma(s,v) = a_2(s,v,h) + O_p(h)
\end{align*}
\end{lemma}

\noindent
B. \it Analysis of \rm $\cA_1(t)$

We can write $\cA_1(t)$ as
$$
\cA_1(t) = \int_0^t \int_0^\tau \int_0^u \psi(u|s) \Gamma(s,v)^{-1} \cS(s,v)\, dv\, du \, ds
$$
with
$$
\cS(s,v) = \frac{1}{mh^3} \sumii Y_{Ri\sd}(v) \KhX (X_{Pi}-\bar{X}_{P0}(s,v))  \left[\lambda_0(v|X_{Pi}) - (X_{Pi}-\bar{X}_{P0}(s,v))
\lambda_0^*(v|s)\right]
$$
By Taylor expansion, we can write
$$
\lambda_0(v|X_{Pi})
= \lambda_0(v|s) + \lambda_0^*(v|s)(X_{Pi}-s) + \frsm{1}{2} \lambda_0^{**}(v|s)(X_{Pi}-s)^2 + \cR(s,v,X_{Pi})
$$
in which $|\cR(s,v,x)| \leq \cR^* |s-x|^3$, with
$\cR^* = \sup_{s,v} |\lambda_0^{***}(v|s)|/6$, where $\lambda_0^{**}(v|s)$ and $\lambda_0^{***}(v|s)$
denote, respectively, the second and third partial derivatives of $\lambda_0(v|s)$ with respect to $s$
We then have
\begin{align*}
& \lambda_0(v|X_{Pi}) - (X_{Pi}-\bar{X}_{P0}(s,v))\lambda_0^*(v|s) \\
& \hspace*{24pt} = \left[ \lambda_0(v|s) + \lambda^*(v|s)(\bar{X}_{P0}(s,v)-s) \right]
+ \frsm{1}{2} \lambda_0^{**}(v|s)(X_{Pi}-s)^2 + \cR(s,v,X_{Pi})
\end{align*}
The term in square brackets does not depend on $i$. Since
$$
\sumii Y_{Ri\sd}(v) \KhX  (X_{Pi}-\bar{X}_{P0}(s,v)) = 0
$$
we get
$$
\cS(s,v)
= \frac{1}{mh^3} \sumii Y_{Ri\sd}(v) \KhX (X_{Pi}-\bar{X}_{P0}(s,v))
\left[ \frsm{1}{2} \lambda_0^{**}(v|s)(X_{Pi}-s)^2 + \cR(s,v,X_{Pi}) \right] \\
$$
We can then write $\cS(s,v) = \cS_1(s,v) + \cS_2(s,v)$, where
\begin{align*}
& \cS_1(s,v) = \frsm{1}{2} \lambda_0^{**}(v|s) \left[ A_3(s,v) h
- (\bar{X}_{P0}(s,v) - s) A_2(s,v) \right] \\
& \cS_2(s,v) = \frac{1}{mh^3} \sumii Y_{Ri\sd}(v) \KhX (X_{Pi}-\bar{X}_{P0}(s,v)) \cR(s,v,X_{Pi})
\end{align*}
We have
$$
\cS_2(s,v) \leq h^2 A_4(s,v) + \cR^* h |\bar{X}_{P0}(s,v)-s| \left[\frac{1}{mh} \sumii Y_{Ri\sd}(v) \KhX
\left| \frac{X_{Pi}-s}{h} \right|^3 \right]
$$
We now consider separately the case of $s \in \cI$ and $s \in \cU$. For $s \in \cI$,
the results of Lemmas 1 and 2 imply that $\cS_1(s,v)=O_p(h^2)$ and $\cS_2(s,v)=O_p(h^2)$,
so that $\cS(s,v)=O_p(h^2)$ and
$$
\int_{[0,t] \cap \cI} \int_0^\tau \int_0^u \psi(u|s) \Gamma(s,v)^{-1} \cS(s,v) dv \, du \, ds = O(h^2)
$$
For $s \in \cU$, the results of Lemmas A1 and A2 imply that $\cS_1(s,v)=O_p(h)$ and $\cS_2(s,v)=O_p(h)$,
so that $\cS(s,v)=O_p(h)$ and
$$
\int_{[0,t] \cap \cU}  \int_0^\tau \int_0^u \psi(u|s) \Gamma(s,v)^{-1} \cS(s,v) dv \, du \, ds = O(h^2)
$$
(recalling that the length of $\cU$ is 2$h$). We thus obtain $\cA_1 = O_p(h^2)$, so that $(mh)^{1/2} \cA_1(t) = o_p(1)$,
since $\nu > 1/4$. \\

\noindent
C. \it Analysis of \rm $\cA_2(t)$

Let $\rho(s,u,v) = \psi(u|s)/\varphi(s,v)$ and define
\begin{align*}
& H(u,v,t,\xi) = \frac{1}{h^2} \int_0^t \frac{\rho(s,u,v)}{\eta_2(s,h)} \Kh{\xi-s} \left(\frac{\xi-s}{h}\right) ds \\
& H_1(u,v,t,\xi) = \frac{1}{h^2} \int_0^t \psi(u|s) (\Gamma(s,v)^{-1} - a_2(s,v,h)^{-1}) \Kh{\xi-s} \left(\frac{\xi-s}{h}\right) ds \\
& H_2(u,v,t,\xi) = \frac{1}{h^2} \int_0^t \psi(u|s) \Gamma(s,v)^{-1} \Kh{\xi-s} \left(\frac{s-\bar{X}_{P0}(s,v)}{h}\right) ds
\end{align*}
Define further
\begin{align*}
& \zeta_i(t) = \int_0^\tau \int_0^u H(u,v,t,X_{Pi}) dM_{Ri\sd}(v) du  \\
& \Delta_{1i}(t) = \int_0^\tau \int_0^u H_1(u,v,t,X_{Pi}) dM_{Ri\sd}(v) du \\
& \Delta_{2i}(t) = \int_0^\tau \int_0^u H_2(u,v,t,X_{Pi}) dM_{Ri\sd}(v) du
\end{align*}
We can then write $\cA_2(t) = \cB(t) + \cB_1(t) + \cB_2(t)$, where
$$
\mathcal{B}(t) =  \frac{1}{m} \sumii \zeta_i(t), \quad
\mathcal{B}_1(t) =  \frac{1}{m} \sumii \Delta_{1i}(t), \quad
\mathcal{B}_2(t) =  \frac{1}{m} \sumii \Delta_{2i}(t)
$$
Our claim is that $(mh)^{1/2} \mathcal{B}(t)$ is asymptotically mean-zero normal, and that $(mh)^{1/2} \cB_1(t)$ and
$(mh)^{1/2} \cB_2(t)$ are both $o_p(1)$.

By (\ref{mz}), $E[\zeta_i(t)]=0$. Thus, $\mathcal{B}$ is the sum of i.i.d.\ mean-zero terms. Accordingly, to show that
$(mh)^{1/2} \, \mathcal{B}(t)$ is asymptotically mean-zero normal we need to show that $h \, \Var(\zeta_i(t))$ converges
to a limit $\kappa(t)$ and that $\zeta_i(t)$ satisfies the Lindeberg condition
\begin{equation}
s_m(t)^{-2} \sumii E\left[\zeta_i(t)^2 I\left(\left|\frac{\zeta_i(t)}{s_m(t)}\right|>\epsilon\right)\right] \rightarrow 0
\quad \mbox{for all } \epsilon > 0
\label{lind}
\end{equation}
where
$$
s_m^2(t) = \Var\left(\sum_{i=1}^m \zeta_i(t) \right) = m \left[ \frac{\kappa(t)}{h} + O(1) \right].
$$
The appended supplementary document provides a proof of the above two assertions, along with a proof that
$(mh)^{1/2} \cB_1(t)$ is $o_p(1)$. The proof of the corresponding result for $\cB_2(t)$ is similar.

\begin{titlepage}
\vspace*{0.125in}
\begin{center}
\LARGE
{\bf Supplemental Web Materials} \\
\vspace*{1em}
\bf An improved fully nonparametric estimator \linebreak[4]
of the marginal survival function \linebreak[4]
based on case-control clustered data
\rm
\normalsize
\end{center}
\vspace{1em}
\begin{center}
\bf David M. Zucker \\
\rm Department of Statistics and Data Science, \\ The Hebrew University of Jerusalem, \\
Mount Scopus, Jerusalem, Israel \\
\it email\rm: david.zucker@mail.huji.ac.il \\
\bf and \\
\bf Malka Gorfine \\
\rm Department of Statistics and Operations Research \\
Tel Aviv University, Ramat Gan, Israel \\
\it email\rm: gorfinem@post.tau.ac.il \\
\vspace*{1in}
\today
\end{center}
\end{titlepage}

\noindent
\bf A. \underline{Introduction and Technical Assumptions for Asymptotic Results} \rm

This document provides details of the proof of the asymptotic properties of the estimator. Notation given in the main paper (including the Appendix)
will be used throughout
without repeating the definitions; additional notation will be defined as needed. In the development below, the notations $O$ and $o$, and similarly $O_p$ and $o_p$, should be understood as being uniform in the relevant arguments.

Below are the technical conditions assumed in deriving the asymptotic results.

1. The kernel $K$ is symmetric, equal to zero outside of $[-1,1]$, and equal to a polynomial inside $[-1,1]$. In addition, $K$ is twice differentiable with bounded derivatives over the entire real line, including the points $-1$ and 1.

2. The bandwidth $h=h_m$ is given by $h_m = \alpha_m m^{-\nu}$, where $1/4 < \nu < 1$ and $\alpha_m \rightarrow \alpha > 0$.

3. We have $g_{min} = \inf_{x \in [0,\tau_0]} g(x) > 0$
and $y_{min} = \inf_{s \in [0, \tau_0]} y(s,\tau) > 0$.

4. The first and second partial derivatives $\dot{\varphi}(s,v)$ and $\ddot{\varphi}(s,v)$
of $\varphi(s,v)$ with respect to $s$ exist and are bounded uniformly over $s$ and $v$. Note that Assumptions 3 and 4 imply
that $\varphi_{min} = \inf_{s \in [0, \tau_0]} \inf_{v \in [0, \tau]} \varphi(s,v) > 0$.

5. The first and second partial derivatives of $\psi(u|s)$ with respect to $s$ exist and are bounded uniformly over $s$ and $v$.

6. The first three partial derivatives of $\lambda_0(v|s)$ with respect to $s$ exist and are bounded uniformly over $s$ and $v$.

\vs
\noindent
\bf B. \underline{Proofs of Lemmas A1 and A2} \rm

We begin with an expanded statement of Lemma A1, continue with the proof of this lemma, and then present Lemma A2 and its proof.

\em{Lemma A1}\rm: For $k$ even we have
$$
E[A_k(s,v,h)] = \left\{
\begin{array}{ll}
a_k(s,v,h) + O(h^2) & s \in \cI \\
a_k(s,v,h) + O(h) & s \in \cU
\end{array} \right.
$$
and for $k$ odd we have
$$
E[A_k(s,v,h)] = \left\{
\begin{array}{ll}
O(h) & s \in \cI \\
a_k(s,v,h) + O(h) & s \in \cU
\end{array} \right.
$$
In addition, for any $k$,
$$
\sup_{s,v} |A_k(s,v,h) - E[A_k(s,v,h)]| = O_p\left((mh)^{-1/2} (\log m)^{1/2}\right)
$$
In general, we have
$$
\sup_{s,v} |A_k(s,v,h) - a_k(s,v,h)| \leq
\sup_{s,v} |A_k(s,v,h) - E[A_k(s,v,h)]| + \sup_{s,v} |E[A_k(s,v,h)] - a_k(s,v,h)|
$$
When $k$ is even, the first term dominates for $s \in I$ while the second term dominates for $s \in \cU$, so that we obtain
$$
\sup_{s,v} |A_k(s,v,h) - a_k(s,v,h)|
= \left\{
\begin{array}{ll}
O_p(m^{-(1-\nu)/2} \, (\log m)^{1/2}) & s \in \cI \\
O_p(m^{-\nu}) & s \in \cU
\end{array}
\right.
$$
When $k$ is odd, we get
$$
\begin{array}{ll}
\sup_{s,v} |A_k(s,v,h)| = O_p(h) & s \in \cI \\
\sup_{s,v} |A_k(s,v,h) - a_k(s,v,h)| =  O_p(h) & s \in \cU
\end{array}
$$

\em{Proof}\rm: The analysis of $E[A_k(s,v)]$ involves a combination of a conditioning argument with the usual change of variable $+$ Taylor
expansion argument. We have
\begin{align*}
E[A_k(s,v,h)] & = E\left[\frac{1}{h}\left(\frac{X_{Pi}-s}{h}\right)^k  \KhX Y_{Ri\sd}(v)\right] \\
& =  E\left[\frac{1}{h} \left(\frac{X_{Pi}-s}{h}\right)^k  \KhX E[Y_{Ri\sd}(v)|X_{Pi}, \delta_{Pi}=0]\right] \\
& = \frac{1}{h} \int_0^{\tau_0} \left(\frac{x-s}{h}\right)^k \Kh{x-s} y(x,v) g(x) dx \\
& = \int_{-s/h}^{(\tau_0-s)/h} r^k K(r) \varphi(s+hr,v) dr
\end{align*}
By Taylor's theorem, we have
$$
\varphi(s+hr,v) = \varphi(s,v) + \dot{\varphi}(s,v)(hr) + \frsm{1}{2} R(s+hr,v)(hr)^2
$$
where $|R(s+hr,v)| \leq \sup_{s,v} |\ddot{\varphi}(s,v)| < \infty$. So we get
\begin{align*}
E[A_k(s,v,h)] & =  \varphi(s,v) \int_{-s/h}^{(\tau_0-s)/h} r^k K(r) dr  \\
& \hspace*{30pt}+ h \dot{\varphi}(s,v)  \int_{-s/h}^{(\tau_0-s)/h} r^{k+1} K(r) dr \\
& \hspace*{30pt} + \frsm{1}{2} h^2  \int_{-s/h}^{(\tau_0-s)/h} r^{k+2} K(r)  R(s+hr,v) dr
\end{align*}
and the claimed result follows.

We now turn to the analysis of $\sup_{s,v} |A_k(s,v,h) - E[A_k(s,v,h)]|$. We use Corollary 2.2 of Gin\'{e} and Guillou (2002),
a result concerning empirical processes.
For $\bx = (x_1, \ldots, x_J) \in \real^J$, define $L_v(\bx) = \sum_{j=1}^J I(x_j \geq v)$ and
$$
\Upsilon_{s,v,h}(x_0,\bx) = Z_k\left(\frac{x_0-s}{h}\right)L_v(\bx)
$$
We can then write
$$
A_k(s,v,h) = \fmh \sumi \Upsilon_{s,v,h}(X_{Pi},X_{R1},\ldots,X_{RJ})
$$
Since $K$ is assumed polynomial on $[-1,1]$, the function $Z_k(r)$ satisfies Gin\'{e} and Guillou's Condition (K$_1$).
Hence, by the arguments in Gin\'{e} and Guillou, the class
$$
\left\{Z_k\left(\frac{\cdot-s}{h}\right): s \in \real, h > 0 \right\}
$$
is a bounded, measurable Vapnik–Chervonenkis (VC) class of functions on $\real$.
Any set of the form $L_v(\bx)=j$ can be expressed as the result of Boolean operations on half-spaces, and hence
the class of sets $\{ \{\bx: L_v(\bx)=j\}, v \in \real, j \in \{0, \ldots, J\} \}$ is a VC class (see Dudley, 1999,
p.\ 141) (this is well known). Further, any set of the form $\{(x_0,\bx): \Upsilon_{s,v,h}(x_0,\bx) \leq b \}$ with $b<0$ can
be expressed as
$$
\bigcup_{j=1}^J \left(\{x_0: Z_k\left(\frac{\cdot-s}{h}\right) \leq b/j \} \times \{\bx: L_v(\bx)=j\} \right)
$$
and any set of this form with $b \geq 0$ can be expressed as
$$
\left[ \bigcup_{j=1}^J \left(\{x_0: Z_k\left(\frac{\cdot-s}{h}\right) \leq b/j \} \times \{\bx: L_v(\bx)=j\} \right) \right]
\cup (\real \times \{\bx: L_v(\bx)=0\})
$$
Recalling that the Cartesian product preserves the VC property, we can conclude that the class of functions
$\Upsilon^*=\{\Upsilon_{s,v,h}: s \in [0,\tau_0], v \in [0, \tau], h > 0 \}$ is a bounded VC class.
Moreover, since the map $(s,v,h,x_0,\bx) \mapsto \Upsilon_{s,v,h}(x_0,\bx)$ is jointly measurable,
the class $\Upsilon^*$ is measurable (see Gin\'{e} and Guillou, bottom of p.\ 911 to top of p.\ 912).
This allows us to apply Gin\'{e} and Guillou's Corollary 2.2.

We have $\sup |\Upsilon_{s,v,h}(x_0,\bx)| \leq U$ with $U = J \sup_r |r|^k K(r)$. Also, a simple standard
calculation shows that $\Var(\Upsilon_{s,v,h}(X_{Pi},X_{R1},\ldots,X_{RJ})) \leq Rh$ for a constant $R$.
Writing $\sigma^2 = Rh$ and letting $\mathfrak{C}$ denote the constant $C$ in  Gin\'{e} and Guillou
Eqn.\ (2.6), we find, after some simple algebra, that for $m$ sufficiently large
$$
\fC \sqrt{m} \sigma \sqrt{\log \frac{U}{\sigma}}
\leq \rho \sqrt{m h \log m}
$$
with $\rho = \fC \sqrt{2R\nu}$.
Thus, writing $\cE_{s,v,h} = E[\Upsilon_{s,v,h}(X_{Pi},X_{R1},\ldots,X_{RJ})]$
and applying Gin\'{e} and Guillou's Corollary 2.2, for $m$ sufficiently large we have
\begin{align*}
& \Pr \left( \left(\frac{mh}{\log m}\right)^{1/2} \sup_{s,v} |A_k(s,v) - E[A_k(s,v)]| > \rho \right) \\
& \hspace*{36pt} = \Pr \left( \left| \sumi \Upsilon_{s,v,h}(X_{Pi},X_{R1},\ldots,X_{RJ}) - \cE_{s,v,h} \right| > \rho \sqrt{mh \log m} \right) \\
&  \hspace*{36pt} \leq \Pr \left( \left| \sumi \Upsilon_{s,v,h}(X_{Pi},X_{R1},\ldots,X_{RJ}) - \cE_{s,v,h} \right|
> \fC \sqrt{m} \sigma \sqrt{\log \frac{U}{\sigma}} \right) \\
&  \hspace*{36pt}\leq \fL_1 \exp\left(- \fL_2 \frac{U}{\sigma} \right) \\
&  \hspace*{36pt} = \fL_1 \exp(- \fL_2 [ \log U/R - \log \alpha_m + \nu \log m ]) \rightarrow 0
\end{align*}
where $\fL_1$ and $\fL_2$ are universal constants. This proves that
$$
\sup_{s,v} |A_k(s,v,h) - E[A_k(s,v,h)]| = O_p\left((mh)^{-1/2} (\log m)^{1/2}\right)
$$

\noindent
\em{Lemma A2}\rm: For $s \in \cI$ we have
\begin{align*}
& \bar{X}_{P0}(s,v) - s = O_p(h^2) \\
& \Gamma(s,v) = a_2(s,v,h) + O_p(m^{-(1-\nu)/2}  \, (\log m)^{1/2})
\end{align*}
and for $s \in \cU$ we have
\begin{align*}
& \bar{X}_{P0}(s,v) - s = \left(\frac{a_1(s,v,h)}{a_0(s,v,h)}\right)h + O_p(h^2) \\
& \Gamma(s,v) = a_2(s,v,h) + O_p(h)
\end{align*}

\em{Proof\rm: Simple algebra yields
\begin{align*}
\bar{X}_{P0}(s,v) - s & = A_0(s,v)^{-1}A_1(s,v)h \\
\Gamma(s,v) & = A_2(s,v) - A_0(s,v) \left( \frac{\bar{X}_{P0}(s,v)-s}{h}\right)^2 \\
& = A_2(s,v) - A_0(s,v)^{-1}(A_1(s,v))^2
\end{align*}
The result now follows immediately from Lemma 1.

\newpage

\noindent
\bf C. \underline{Analysis of $\cA_2(t)$} \rm

As stated in the Appendix of the main paper, we let $\rho(s,u,v) = \psi(u|s)/\varphi(s,v)$ and define
\begin{align*}
& H(u,v,t,\xi) = \frac{1}{h^2} \int_0^t \frac{\rho(s,u,v)}{\eta_2(s,h)} \Kh{\xi-s} \left(\frac{\xi-s}{h}\right) ds \\
& H_1(u,v,t,\xi) = \frac{1}{h^2} \int_0^t \psi(u|s) (\Gamma(s,v)^{-1} - a_2(s,v,h)^{-1}) \Kh{\xi-s} \left(\frac{\xi-s}{h}\right) ds \\
& H_2(u,v,t,\xi) = \frac{1}{h^2} \int_0^t \psi(u|s) \Gamma(s,v)^{-1} \Kh{\xi-s} \left(\frac{s-\bar{X}_{P0}(s,v)}{h}\right) ds
\end{align*}
In addition, we define
\begin{align*}
& \zeta_i(t) = \int_0^\tau \int_0^u H(u,v,t,X_{Pi}) dM_{Ri\sd}(v) du \\
& \Delta_{1i}(t) = \int_0^\tau \int_0^u H_1(u,v,t,X_{Pi}) dM_{Ri\sd}(v) du \\
& \Delta_{2i}(t) = \int_0^\tau  \int_0^u H_2(u,v,t,X_{Pi}) dM_{Ri\sd}(v) du
\end{align*}
We can then write $\cA_2(t) = \cB(t) + \cB_1(t) + \cB_2(t)$, where
$$
\mathcal{B}(t) =  \frac{1}{m}  \sumi \zeta_i(t), \quad
\mathcal{B}_1(t) =  \frac{1}{m}  \sumi\Delta_{1i}(t), \quad
\mathcal{B}_2(t) =  \frac{1}{m} \sumi\Delta_{2i}(t)
$$
Our claim is that $\sqrt{mh} \, \mathcal{B}(t)$ is asymptotically mean-zero normal, and that $\sqrt{mh} \, \cB_1(t)$ and
$\sqrt{mh} \, \cB_2(t)$ are both $o_p(1)$.

As noted in the Appendix of the main paper, $\mathcal{B}(t)$ is the sum of i.i.d.\ mean-zero terms. Accordingly, to show that
$\sqrt{mh} \, \mathcal{B}(t)$ is asymptotically mean-zero normal we need to show that $h \Var(\zeta_i(t))$ converges
to a limit $\kappa(t)$ and that $\zeta_i(t)$ satisfies the Lindeberg condition
\begin{equation}
s_m(t)^{-2} \sumi E\left[\zeta_i(t)^2 I\left(\left|\frac{\zeta_i(t)}{s_m(t)}\right|>\epsilon\right)\right] \rightarrow 0 \quad \forall \epsilon > 0
\label{lind}
\end{equation}
where
$$
s_m(t)^2 = \Var\left(\sum_{i=1}^m \zeta_i(t)\right) = m \left[ \frac{\kappa(t)}{h} + O(1) \right]
$$

\noindent
1. \underline{Analysis of $\Var(\zeta_i(t))$}

We can write $H(u,v,t,\xi)$ as
$$
H(u,v,t,\xi) = \frac{1}{h} \int_{-1}^1 rK(r) I\left( r \in \left[\frac{\xi-t}{h},\frac{\xi}{h}\right]\right)
\frac{\rho(\xi-hr,u,v)}{\eta_2(\xi-hr,h)} dr
$$
The relevant range of $\xi$ is $[0,\tau_0]$. The analysis of $H(u,v,t,\xi)$ divides into several cases.
To ease the presentation, we assume that $t < \tau_0$. A similar development can be given for $t=\tau_0$.

Case 1, $\xi = \omega h$ with $\omega \in [0,1]$: In this we case we have $H(u,v,t,\xi) = -h^{-1} \cP_1(\omega) \rho(0,u,v) + O(1)$, where
$$
\cP_1(\omega) = \int_{-\omega}^{1-\omega} \frac{rK(r)}{\mu_2(\omega+r)} \, dr - \frac{\mu_1(-1+\omega)}{\mu_2(1)}
$$

Case 2, $\xi = (1+\omega)h$ with $\omega \in [0,1]$: In this we case we have $H(u,v,t,\xi) = -h^{-1} \cP_2(\omega) \rho(0,u,v) + O(1)$, where
$$
\cP_2(\omega) = \int_{-1}^{-\omega} \frac{rK(r)}{\mu_2(1+\omega+r)} \, dr - \frac{\mu_1(\omega)}{\mu_2(1)}
$$

Case 3, $\xi \in [2h,t-h]$: In this case the indicator equals 1 and $\xi-hr \in \cI$ for all $r \in [-1,1]$, and hence, recalling that
$\mu_1(1)=0$, we get
$H(u,v,t,\xi) = -\dot{\rho}(\xi,u,v) + O(h)$, where $\dot{\rho}(s,u,v)$ is the partial derivative of
$\rho(s,u,v)$ with respect to $s$.

Case 4, $\xi = t + \omega h$ with $\omega \in [-1,1]$: In this case, $H(u,v,t,\xi) = -h^{-1} \mu_1(-\omega) \rho(t,u,v)/\mu_2(1) + O(1)$.

Case 5, $\xi > t+h$: In this case the indicator equals 0 for all $r \in [-1,1]$ and so $H(u,v,t,\xi)=0$.

Define
\begin{align*}
& \cV(\xi,t)
= E\left[\left. \int_0^\tau \left(\int_0^u H(u,v,t,\xi) dM_{i\sd}(v) du \right)^2 \right| X_{Pi}=\xi, \delta_{Pi}=0 \right] \\
& \cV^*(\xi,\xi^\pr)
= E\left[\left. \left(\int_0^\tau \int_0^u \rho(\xi^\pr,u,v) dM_{i\sd}(v) du \right)^2 \right| X_{Pi}=\xi, \delta_{Pi}=0 \right] \\
& \dot{\cV}^*(\xi)
= E\left[\left. \left(\int_0^\tau \int_0^u \dot{\rho}(\xi,u,v) dM_{i\sd}(v) du \right)^2 \right| X_{Pi}=\xi, \delta_{Pi}=0 \right]
\end{align*}
We then have
$$
\cV(\xi,t) = \left\{
\begin{array}{ll}
$$
h^{-2} \cV^*(\xi,0)\cP_1(\xi/h)^2 + O(h^{-1}) & \xi \in [0,h] \\
h^{-2} \cV^*(\xi,0)\cP_2(\xi/h-1)^2 + O(h^{-1}) & \xi \in [h,2h] \\
\dot{\cV}^*(\xi) + O(h) & \xi \in [2h,t-h]  \\
h^{-2} \cV^*(\xi,t)(\mu_1(-(\xi-t)/h)/\mu_2(1))^2 + O(h^{-1}) & \xi \in [t-h,t+h] \\
0 & \xi > t + h
\end{array}
\right.
$$
Accordingly,
$$
\Var(\zeta_i(t)) = E[\zeta_i(t)^2]
= \int_0^{\tau_0} g(\xi) \cV(\xi) d\xi = \cC_1 + \cC_2 + \cC_3 + \cC_4
$$
where
\begin{align*}
& \cC_1
= \int_0^h g(\xi) \cV(\xi,0,t) d\xi
= h \int_0^1 g(\omega h) \cV(\omega h,0,t) d\omega
= h^{-1} g(0) \cV^*(0,0) \int_0^1 \cP_1(\omega)^2 d\omega  + O(1) \\
& \cC_2
= \int_h^{2h} g(\xi) \cV(\xi,t) d\xi
= h \int_1^2 g(\omega h) \cV(\omega h,0,t) d\omega
= h^{-1} g(0) \cV^*(0,0) \int_0^1 \cP_2(\omega)^2 d\omega  + O(1) \\
& \cC_3 = \int_{2h}^{t-h} g(\xi) \cV(\xi,t) d\xi
= \int_{2h}^{t-h} g(\xi) \dot{\cV}^*(\xi) d\xi + O(h) \\
& \cC_4 = \int_{t-h}^{t+h} g(\xi) \cV(\xi,t) d\xi
= h^{-1}g(t) \cV^*(t,t) \mu_2(1)^{-2} \int_{-1}^1 \mu_1(\omega)^2 d\omega + O(1)
\end{align*}
In other words,
$$
\Var(\zeta_i(t)) = \frac{\kappa(t)}{h} + O(1)
$$
where
$$
\kappa(t) = g(0) \cV^*(0,0) \left[ \int_0^1 \cP_1(\omega)^2 d\omega
+ \int_0^1 \cP_2(\omega)^2 d\omega \right]
+ g(t) \cV^*(t,t) \mu_2(1)^{-2} \int_{-1}^1 \mu_1(\omega)^2 d\omega
$$

\noindent
2. \underline{Proof of Lindeberg Condition}

We have
\begin{align*}
|\zeta_i(t)| & = \left| \int_0^\tau \int_0^u H(u,v,t,X_{Pi}) dM_{Ri\sd}(v) du \right| \\
& \leq \int_0^\tau \int_0^u |H(u,v,t,X_{Pi})| dN_{Ri\sd}(v) du
+ \int_0^\tau \int_0^u |H(u,v,t,X_{Pi})| Y_{Ri\sd}(v) \lambda(v) dv du \\
& \leq (1+\lambda_{max})J \tau \sup_{u,v,t,\xi} |H(u,v,t,\xi)| \\
& \leq \left[(1+\lambda_{max})J \tau \right] \left[\int_{-1}^1 |r| K(r) dr\right]
\mu_2(0)^{-1} \sup_{s,u,v} |\rho(s,u,v)| h^{-1} = \cM h^{-1}
\end{align*}
with $\cM$ defined in the obvious manner. Thus,
$$
\left|\frac{\zeta_i(t)}{s_m(t)}\right| \leq \frac{\cM}{[(\kappa(t)+O(h))mh]^{1/2}} \rightarrow 0
$$
Thus, the Lindeberg condition (\ref{lind})  is satisfied.

\noindent
3. \underline{Analysis of $\mathcal{B}_1(t)$ and $\mathcal{B}_2(t)$}

We show here that $\sqrt{mh} \, \, \cB_1(t) = o_p(1)$; the argument for $\sqrt{mh} \, \, \cB_2(t)$ is similar.
For simplicity of exposition, we present the proof for the case $J=2$.
Define the filtration
$\mathcal{F}_{v} = \sigma(\mathcal{F}_0, \{N_{Rij}(d), Y_{Rij}(d);$ $i = 1, \ldots, m; j = 1, 2; \, d \in [0,v]\})$.
We can write
\begin{align*}
& \lim_{d \downarrow 0} \Pr(N_{Ri1}(t+d)-N_{Ri1}(t)=1|\cF_{v-}) \\
& \hspace*{18pt} = Y_{Ri1}(v) \left[ Y_{Ri2}(v) \lambda_0(v|v) + (1-Y_{Ri2}(v))(1-N_{Ri2}(v))\lambda_0(v|X_{Ri2}) + N_{Ri2}(v) \lambda_1(v|X_{Ri2}) \right]
\end{align*}
A similar equality holds for $\lim_{d \downarrow 0} \Pr(N_{Ri2}(t+d)-N_{Ri2}(t)=1|\cF_{v-})$.
So if we define
\begin{align*}
\tilde{\lambda}_i(v) & =
Y_{Ri1}(v) \left[ Y_{Ri2}(v) \lambda_0(v|v) + (1-Y_{Ri2}(v))(1-N_{Ri2}(v))\lambda_0(v|X_{Ri2}) + N_{Ri2}(v) \lambda_1(v|X_{Ri2}) \right] \\
& \hspace*{8pt} + Y_{Ri2}(v) \left[ Y_{Ri1}(v) \lambda_0(v|v) + (1-Y_{Ri1}(v))(1-N_{Ri1}(v))\lambda_0(v|X_{Ri1}) + N_{Ri1}(v) \lambda_1(v|X_{Ri1}) \right]
\end{align*}
then the process $\tilde{M}_{R i \sd}(v)$ defined by $d\tilde{M}_{R i \sd}(v) = \sum_{i=1}^m (dN_{Ri\sd}(v)- \tilde{\lambda}_i(v) dv)$
is a martingale with respect to the filtration $\cF_v$.

Define
$$
\mathcal{B}_1(t,u) =  \frac{1}{m}  \sumi\Delta_{1i}(t,u)
$$
with
$$
\Delta_{1i}(t,u) = \int_0^u H_1(u,v,t,X_{Pi}) dM_{Ri\sd}(v)
$$
We can write $\sqrt{mh} \, \, \cB_1(t,u) = \cB_{1}^*(t,u,u) + \cB_1^{**}(t,u)$, where
\begin{align*}
\cB_{1}^*(t,u,w) & = \sqrt{mh} \left[ \frac{1}{m} \sumi \int_0^u H_1(w,v,t,X_{Pi}) d\tilde{M}_{Ri\sd}(v) \right] \\
\cB_{1}^{**}(t,u) & =  \sqrt{mh} \left[ \frac{1}{m} \sumi \int_0^u H_1(w,v,t,X_{Pi}) (d{M}_{Ri\sd}(v) - d\tilde{M}_{Ri\sd}(v)) \right] \\
& = \sqrt{mh} \left[ \frac{1}{m} \sumi \int_0^u H_1(w,v,t,X_{Pi}) (\tilde{\lambda}_i(v) - Y_{Ri\sd}(v) \lambda_0(v|X_{Pi})) dv \right]
\end{align*}
Note that $E[\tilde{\lambda}_i(v) - Y_{Ri\sd}(v) \lambda_0(v|X_{Pi})]=0$.

Now, since  $H_1(u,v,t,X_{Pi})$, viewed as a process in $v$, is predictable with respect to $\cF_v$,
the process $\cB_{1}^*(t,u,w)$ viewed as a process in $u$, is a martingale with respect to $\cF_v$, with predictable
variation process given by
\begin{equation}
\langle \cB_1^*(t,\cdot,w),\cB_1^*(t,\cdot,w) \rangle (u)
= h \int_0^u \left[ \frac{1}{m} \sumi H_1(w,v,t,X_{Pi})^2  \tilde{\lambda}_i(v) \right] dv
\label{pvar}
\end{equation}
We can write $H_1(w,v,t,\xi) = H_{1a}(w,v,t,\xi) +  H_{1b}(w,v,t,\xi)$ with
\begin{align*}
H_{1a}(w,v,t,\xi) = \frac{1}{h^2} \int_{[0,t] \cap \cI} \psi(w|s) (\Gamma(s,v)^{-1} - a_2(s,v,h)^{-1}) \Kh{\xi-s} \left(\frac{\xi-s}{h}\right) ds \\
H_{1b}(w,v,t,\xi) = \frac{1}{h^2} \int_{[0,t] \cap \cI} \psi(w|s) (\Gamma(s,v)^{-1} - a_2(s,v,h)^{-1}) \Kh{\xi-s} \left(\frac{\xi-s}{h}\right) ds
\end{align*}
Now,
$$
H_{1a}(w,v,t,\xi)
\leq \sup_{s \in \cI, v \in [0,\tau]} |\Gamma(s,v)^{-1}-a_2(s,v)^{-1}| \sup_{w,s} |\psi(w|s)| \fA(\xi)
$$
with
$$
\fA(\xi) = \frac{1}{h^2} \int_0^t \Kh{\xi-s} \left|\frac{\xi-s}{h}\right| ds
$$
Defining
$$
\fB(r) = \int_{-1}^r |r^\pr| K(r^\pr) dr^\pr
$$
we have $\fA(\xi) \leq \fA^\pr(\xi)$ with
$$
\fA^\pr(\xi) = h^{-1} \fB \left(\frac{t-\xi}{h}\right)
$$
and
$$
E[\fA^\pr(X_{Pi})]
= \int_0^{\tau_0}  h^{-1} \fB \left(\frac{t-\xi}{h}\right) g(\xi) d\xi
\leq \int_{-t/h}^{(\tau_0-t)/h} \fB(\xi^\pr) g(t+ h\xi^\pr) d\xi^\pr
\leq \fB(1)
$$
Thus $\fA(X_{Pi}) = O_p(1)$.
Hence, using the result of Lemma A2, we obtain
$$
|H_{1a}(w,v,t,X_{Pi})|
= O_p(m^{-(1-\nu)/2} (\log m)^{1/2}) = O_p(1)
$$
since $\nu < 1$.
Similarly, again using the result of Lemma A2, we find that
$|H_{1b}(w,v,t,X_{Pi})|= O_p(1)$.
Accordingly, the term in brackets in (\ref{pvar}) is $O_p(1)$.
It follows from Lenglart's inequality (see, e.g., Andersen and Gill, 1982, Thm.\ I.1(b)) that
for any given $w$ (and in particular for $w=u$), $\sup_{u \in [0,\tau]} |\cB_{1}^*(t,u,w)| = O_p(\sqrt{h})$.

In regard to $\cB_{1}^{**}(t,u)$, using the Cauchy-Schwarz inequality, we have
\begin{align*}
|\cB_{1}^{**}(t,u)| & = \sqrt{mh} \left| \int_0^u
\int_0^t \psi(u|s) (\Gamma(s,v)^{-1} - a_2(s,v,h)^{-1}) \right. \\
& \hspace*{60pt} \left. \left[ \frac{1}{mh^2} \sumi \Kh{X_{Pi}-s} \left(\frac{X_{Pi}-s}{h}\right)
(\tilde{\lambda}_i(v) - Y_{Ri\sd}(v) \lambda_0(v|X_{Pi})) \right] ds \, dv \right| \\
& \leq \sqrt{mh} \sqrt{\cQ_1 \cQ_2}
\end{align*}
with
\begin{align*}
\cQ_1 & = \int_0^u \int_0^t \psi(u|s) (\Gamma(s,v)^{-1} - a_2(s,v,h)^{-1})^2 ds \, dv \\
\cQ_2 & = \int_0^u \int_0^t \psi(u|s)
\left[ \frac{1}{mh^2} \sumi \Kh{X_{Pi}-s} \left(\frac{X_{Pi}-s}{h}\right)
(\tilde{\lambda}_i(v) - Y_{Ri\sd}(v) \lambda_0(v|X_{Pi})) \right]^2 ds \, dv
\end{align*}
Now,
\begin{align*}
E[\cQ_2] & = \int_0^u \int_0^t \psi(u|s) E \left[ \left\{\frac{1}{mh^2} \sumi \Kh{X_{Pi}-s} \left(\frac{X_{Pi}-s}{h}\right)
(\tilde{\lambda}_i(v) - Y_{Ri\sd}(v) \lambda_0(v|X_{Pi})) \right\}^2  \right] ds \, dv\\
& = (mh^2)^{-1} \int_0^u \int_0^t \psi(u|s)
E \left[\frac{1}{h^2} K^2\left(\frac{X_{Pi}-s}{h}\right) \left(\frac{X_{Pi}-s}{h}\right)^2
(\tilde{\lambda}_i(v) - Y_{Ri\sd}(v) \lambda_0(v|X_{Pi}))^2 \right] ds \, dv \\
& = (mh^2)^{-1} \int_0^u \int_0^t \int_0^{\tau_0}
h^{-2} \psi(u|s) K^2\left(\frac{\xi-s}{h}\right) \left(\frac{\xi-s}{h}\right)^2
(\tilde{\lambda}_i(v) - Y_{Ri\sd}(v) \lambda_0(v|\xi))^2 g(\xi) d\xi \,ds \, dv \\
& = O((mh^2)^{-1})
\end{align*}
so that $\cQ_2 =  O_p((mh^2)^{-1})$.
In addition, by Lemma A2, $\cQ_1 =  O_p(m^{-(1-\nu)}  \, (\log m))$.
We thus find that $\cB_{1}^{**}(t,u)=O_p(m^{-(1-2\nu)/2} (\log m)^{1/2})=o_p(1)$. \\

\noindent
\bf D. \underline{Time Transformation According to the EDF of the Proband Observation Times} \rm

In this section, we sketch the proof that the consistency and asymptotic normality of our estimator is maintained under a time transformation
based on an estimate $G_m$ of the cumulative distribution function $G$ of the proband observation times. In this proof we need to assume that
$G$ has four bounded derivatives and that a smooth estimate of $G$ is used.
We conjecture that the result holds without these conditions
(in our simulations, we obtained good results taking $G_m$ to be a linearly interpolated version of the empirical CDF), but we do not have a proof.

We take
$$
G_m(t) = \frac{1}{m} \sumi \mathcal{K}\left(\frac{t-X_{Pi}}{b_m}\right)
$$
where
$$
\mathcal{K}(a) =  \int_{-\infty}^a K(c) dc
$$
with $b_m$ chosen as described in Schuster (1969) so that $G_m(t)$ and its first three derivatives converge uniformly to $G$ and its first
three derivatives.
We define $D=G^{-1}$ and $D_m=G_m^{-1}$.

For any thrice differentiable inverse CDF function $\cD$ and any $\bar{s} \in [0,1]$, define
\begin{align*}
& \Lambda_\cD(\bs) = \Lambda(\cD(\bs)) \\
& S_\ocd(u|\bs) = S_0(u|\cD(\bs)) \\
& \Lambda_\ocd(u|\bs) = \Lambda_0(u|\cD(\bs)) \\
& \Lambda_\ocd^*(u|\bs) = \frac{\partial}{\partial \bs} \Lambda_\ocd(u|\bs) = \Lambda_0^*(u|\cD(\bs)) \cD^\pr(\bs) \\
& \lambda_\ocd(u|\bs) = \frac{\partial}{\partial u} \Lambda_\ocd(u|\bs) = \lambda_0(u|\cD(\bs)) \\
& \lambda_\ocd^*(u|\bs) = \frac{\partial}{\partial \bs} \lambda_\ocd(u|\bs) \\
& \lambda_\ocd^{**}(u|\bs) = \frac{\partial^2}{\partial \bs^2} \lambda_\ocd(u|\bs) \\
& \lambda_\ocd^{***}(u|\bs) = \frac{\partial^3}{\partial \bs^3} \lambda_\ocd(u|\bs) \\
& \psi_\cD(u,\bs) = \psi(u,\cD(\bs))
\end{align*}
Also, for any distribution function $\cG$, define
\begin{align*}
A_k(\bs,v,h,\cG) & =
\frac{1}{mh} \sum_{i=1}^m Y_{Ri\sd}(v) Z_k\left(\frac{\cG(X_{Pi})-\bs}{h}\right) \\
\cM(\bs,v,\cG) & = A_1(\bs,v,h,\cG)/A_0(\bs,v,h,\cG) \\
\Gamma(\bs,v,\cG) & = A_2(\bs,v,h,\cG) - A_0(\bs,v,h,\cG)^{-1}(A_1(\bs,v,h,\cG))^2
\end{align*}
For any given nonrandom $\cG$, the analogues of Lemmas A1 and A2 hold for the above quantities.
Our estimator $\hat{\Lambda}_{0,D}^*(u|\bs)$ of $\Lambda_{0,D}^*(u|\bs)$ is
$$
\hat{\Lambda}_{0,D}^*(u|\bs) = \int_0^u {\Gamma(\bs,v,G_m)}^{-1} \left[\frac{1}{mh^3} \sum_{i=1}^m \Kh{G_m(X_{Pi})-\bs}
(G_m(X_{Pi})-\cM(\bs,v,G_m)) dN_{Ri\sd}(v) \right]
$$
our estimator of $\Lambda_D(\bs)$ is
$$
\hat{\Lambda}_D(\bt) =  -\int_0^{\bt} \int_0^\tau \hat{\psi}_{D}(u,\bs) \hat{\Lambda}_{0,D}^*(u|\bs)  \, du \, d\bs
$$
and our estimator of $\Lambda(t)$ is $\hat{\Lambda}(t)=\hat{\Lambda}_D(G_m(t))$.

In the analysis of $\hat{\Lambda}_D(\bt) - \Lambda_D(\bt)$,
the analogues of $\cA_1(t)$ and $\cA_2(t)$ are $\cA_1(\bt,G_m)$ and $\cA_2(\bt,G_m)$, where
\begin{align*}
& \cA_1(\bt,\cG) = \frac{1}{mh^3} \sum_{i=1}^m \int_0^{\bt} \int_0^\tau \int_0^u \psi_D(u,\bs) \Gamma(\bs,v,\cG)^{-1} Y_{Ri\sd}(v)
\Kh{\cG(X_{Pi})-\bs} (\cG(X_{Pi})-\cM(\bs,v,\cG))  \\
& \hspace*{100pt} (\lambda_{0,\cG^{-1}}(v|\cG(X_{Pi})) - (\cG(X_{Pi})-\cM(\bs,v,\cG))\lambda_{0,D}^*(v|\bs)) dv \, \, du \, d\bs \\
& \cA_2(\bt,\cG) = \frac{1}{mh^3} \sum_{i=1}^m \int_0^{\bt} \int_0^\tau \int_0^u \psi_D(u,\bs)\Gamma(\bs,v,\cG)^{-1}
\Kh{\cG(X_{Pi})-\bs} (\cG(X_{Pi})-\cM(\bs,v,\cG)) dM_{Ri\sd}(v) \, du \, d\bs
\end{align*}
We can write $\cA_1(\bt,\cG) = \cA_{1a}(\bt,\cG) - \cA_{1b}(\bt,\cG)$, where
\begin{align*}
& \cA_{1a}(\bt,\cG) = \frac{1}{mh^3} \sum_{i=1}^m \int_0^{\bt} \int_0^\tau \int_0^u \psi_D(u,\bs) \Gamma(\bs,v,\cG)^{-1} Y_{Ri\sd}(v)
\Kh{\cG(X_{Pi})-\bs} (\cG(X_{Pi})-\cM(\bs,v,\cG))  \\
& \hspace*{100pt} (\lambda_{0,\cG^{-1}}(v|\cG(X_{Pi})) - (\cG(X_{Pi})-\cM(\bs,v,\cG))\lambda_{0,\cG^{-1}}^*(v|\bs)) dv \, \, du \, d\bs \\
& \cA_{1b}(\bt,\cG) =  \int_0^{\bt} \int_0^\tau \psi_D(u,\bs)
(\Lambda_{0,\cG^{-1}}^*(u|\bs) - \Lambda_{0,D}^*(u|\bs))  du \, d\bs
\end{align*}
By the same Taylor expansion argument as in the Appendix of the main paper, we can write
\begin{align*}
& \cA_{1a}(\bt,\cG) = \frac{1}{mh^3} \sum_{i=1}^m \int_0^{\bt} \int_0^\tau \int_0^u \psi_D(u,\bs) \Gamma(\bs,v,\cG)^{-1} Y_{Ri\sd}(v) \Kh{\cG(X_{Pi})-\bs} \\
& \hspace*{100pt}
(\cG(X_{Pi})-\cM(\bs,v,\cG))
\left[ \frsm{1}{2} \lambda_{0,\cG^{-1}}^{**}(v|\bs)(\cG(X_{Pi})-\bs)^2 + \cR(\bs,v,\cG(X_{Pi})) \right]
\end{align*}
where $|\cR(\bs,v,\bx)| \leq \cR_m^* |\bs-\bx|^3$ with $\cR_m^* = O_p(1)$.
We have $\sqrt{mh} \, \cA_{1a}(\bt,G)$ by the same argument as in the Appendix of the main paper,
which leaves us to deal with $\cA_{1a}(\bt,G_m)-\cA_{1a}(\bt,G)$ and $\cA_{1b}(\bt,G_m)$,
The quantity $\cA_{1a}(\bt,G_m)-\cA_{1a}(\bt,G)$ can be broken up into a series of various terms.
A typical term is
\begin{align*}
\Psi & = \frac{1}{mh^3} \sum_{i=1}^m \int_0^{\bt} \int_0^\tau \int_0^u \psi_D(u,\bs) \Gamma(\bs,v,G)^{-1} Y_{Ri\sd}(v) \left[ \Kh{G_m(X_{Pi})-\bs} - \Kh{G(X_{Pi})-\bs} \right] \\
& \hspace*{100pt}
(G(X_{Pi})-\cM(\bs,v,G))
\left[ \frsm{1}{2} \lambda_{0,D}^{**}(v|\bs)(G(X_{Pi})-\bs)^2 + \cR(\bs,v,G(X_{Pi})) \right] dv \, \, du \, d\bs
\end{align*}
We can write $\Psi=\Psi_1 - \Psi_2$, where
\begin{align*}
\Psi_1 & = \frac{1}{mh^3} \sum_{i=1}^m \int_0^{\bt} \int_0^\tau \int_0^u \psi_D(u,\bs) \Gamma(\bs,v,G)^{-1} Y_{Ri\sd}(v) \left[ \Kh{G_m(X_{Pi})-\bs} - \Kh{G(X_{Pi})-\bs} \right] \\
& \hspace*{100pt}
(G(X_{Pi})-\bs)
\left[ \frsm{1}{2} \lambda_{0,D}^{**}(v|\bs)(G(X_{Pi})-\bs)^2 + \cR(\bs,v,G(X_{Pi})) \right] dv \, \, du \, d\bs \\
\Psi_2 & = \frac{1}{mh^3} \sum_{i=1}^m \int_0^{\bt} \int_0^\tau \int_0^u \psi_D(u,\bs) \Gamma(\bs,v,G)^{-1} Y_{Ri\sd}(v) \left[ \Kh{G_m(X_{Pi})-\bs} - \Kh{G(X_{Pi})-\bs} \right] \\
& \hspace*{100pt}
(\cM(\bs,v,G) - \bs)
\left[ \frsm{1}{2} \lambda_{0,D}^{**}(v|\bs)(G(X_{Pi})-\bs)^2 + \cR(\bs,v,G(X_{Pi})) \right] dv \, \, du \, d\bs
\end{align*}
Write $\Delta_i = (G_m(X_{Pi})-G(X_{Pi}))/h$ and $\Delta = \|G_m - G \|_\infty/h$.
We have
\begin{align*}
|\Psi_1 | & \leq O_p(1) \left[
\frac{1}{m} \sum_{i=1}^m \int_0^{\bt} \left|  \Kh{G_m(X_{Pi})-\bs} - \Kh{G(X_{Pi})-\bs} \right| \right. \\
& \hspace*{100pt} \left. \left\{ \left| \frac{G(X_{Pi}) - \bs}{h} \right|^3
+ \left| \frac{G(X_{Pi}) - \bs}{h} \right|^4 \right\} d\bs \right] \\[1ex]
& = O_p(1) \left[ \frac{h}{m} \sum_{i=1}^m \int_{-G(X_{Pi})/h}^{(\bt-G(X_{Pi}))/h}
\left|  K\left(r + \frac{G_m(X_{Pi})-G(X_{Pi})}{h}\right) - K(r) \right|
(|r|^3 + |r|^4) dr \right] \\
& \leq O_p(1) \left[ \frac{h}{m} \sum_{i=1}^m \int_{-1-|\Delta|}^{1+|\Delta|}
(|r|^3 + |r|^4) dr |K(r + \Delta_i)-K(r)| dr \right] \\
& \leq O_p(1) \left[ \frac{h}{m} \sum_{i=1}^m |\Delta_i| \int_{-1-|\Delta|}^{1+|\Delta|}
(|r|^3 + |r|^4) dr \right] \\
& \leq O_p(1)[h\Delta(1+\Delta)^4]
\end{align*}
Recalling that $\|G_m - G \|_\infty = O_p(m^{-1/2})$, we obtain $|\Psi_1| = O_p(m^{-1/2})$.
Thus $\sqrt{mh} \, \Psi_1 = o_p(1)$.
Similarly, using Lemma A2, $\sqrt{mh} \, \Psi_2 = o_p(1)$.

Next, regarding $\cA_{1b}(\bt,G_m)$, we can write
\begin{align*}
\Lambda_{0,D_m}^*(u|\bs) - \Lambda_{0,D}^*(u|\bs)
& = \Lambda_0^*(u|D_m(\bs)) D_m^\pr(\bs) - \Lambda_0^*(u|D(\bs)) D^\pr(\bs) \\
& = (\Lambda_0^*(u|D_m(\bs))-\Lambda_0^*(u|D(\bs)))D_m^\pr(\bs)
+ \Lambda_0^*(u|D(\bs))(D_m^\pr(\bs) -  D^\pr(\bs)) \\
& = \Lambda_{0,D}^{**}(u|s^*)(D_m(\bs) -  D(\bs)) + \Lambda_0^*(u|D(\bs))(D_m^\pr(\bs) -  D^\pr(\bs))
\end{align*}
with $s^*$ between $D_m(\bs)$ and $D(\bs)$. The first of the above two terms is $O_p(m^{-1/2})$.
Regarding the second term, using integration by parts we can write
\begin{align*}
& \int_0^{\bt} \psi_D(u,\bs) \Lambda_0^*(u|D(\bs))(D_m^\pr(\bs) -  D^\pr(\bs)) d\bs \\
& \hspace*{20pt} = \left[ \psi_D(u,\bs) \Lambda_0^*(u|D(\bs))(D_m(\bs) -  D(\bs)) \right]_0^{\bt}
- \int_0^{\bt} \left[ \frac{\partial}{\partial \bs} \psi_D(u,\bs) \Lambda_0^*(u|D(\bs)) \right](D_m(\bs) -  D(\bs)) d\bs \\
& \hspace*{20pt} = O_p(m^{-1/2})
\end{align*}
Thus $\sqrt{mh} \, \cA_{1b}(\bt,G_m) = o_p(1)$.

We turn now to $\cA_2(\bt,G_m)$. By the same arguments as before we find that $\sqrt{mh} \, \cA_2(\bt,G)$ converges in distribution to a mean-zero normal
distribution. This leaves us to deal with $\cA_2(\bt,G_m) - \cA_2(\bt,G)$. This quantity can be broken up into a series of terms, a typical one of which is
$$
\int_0^\tau \Omega(\bt,u) du
$$
with
\begin{align*}
\Omega(\bt,u) & = \frac{1}{mh^3} \sum_{i=1}^m \int_0^{\bt} \int_0^\tau \int_0^u \psi_D(u,\bs)a_2(\bs,v,G)^{-1}
\left[ \Kh{G_m(X_{Pi})-\bs} - \Kh{G(X_{Pi})-\bs} \right] \\
& \hspace*{120pt} (G(X_{Pi})-\bs) \, dM_{Ri\sd}(v) \, du \, d\bs \\
& = \frac{1}{mh^2} \sum_{i=1}^m \int_0^{\bt} \int_0^\tau \int_0^u \psi_D(u,\bs)a_2(\bs,v,G)^{-1}
\left[ \Kh{G_m(X_{Pi})-\bs} - \Kh{G(X_{Pi})-\bs} \right] \\
& \hspace*{120pt} \left(\frac{G(X_{Pi})-\bs}{h}\right)  \, dM_{Ri\sd}(v) \, du \, d\bs
\end{align*}
This term can be dealt using an argument similar to that used for $\cB_1(t,u)$.
We can write $\Omega(u) = \Omega^*(\bt,u,w) + \Omega^{**}(\bt,u)$ with
\begin{align*}
& \Omega^*(\bt,u,w) = \frac{1}{m} \sum_{i=1}^m \int_0^u \tilde{H}(w,v,\bt,X_{Pi}) \tilde{M}_{Ri}(v) \\
& \Omega^{**}(\bt,u) = \frac{1}{m} \sum_{i=1}^m \int_0^u \tilde{H}(w,v,\bt,X_{Pi}) (\tilde{\lambda}_i(v) - Y_{Ri\sd}(v) \lambda_{0,D}(v|G(X_{Pi}))) dv \\
& \tilde{H}(w,v,t,X_{Pi})
= \frac{1}{h^2} \int_0^{\bt} \psi_D(u,\bs)a_2(\bs,v,G)^{-1} \\
& \hspace*{130pt}\left[ \Kh{G_m(X_{Pi})-\bs} - \Kh{G(X_{Pi})-\bs} \right]
\left(\frac{G(X_{Pi})-\bs}{ h}\right) \, d\bs
\end{align*}
We have
$$
|\tilde{H}(w,v,\bt,X_{Pi})|
= \tilde{H}_a(w,v,\bt,X_{Pi}) + \tilde{H}_b(w,v,\bt,X_{Pi})
$$
with
\begin{align*}
\tilde{H}_a(w,v,\bt,X_{Pi})
& = \frac{1}{h^3} \int_0^{\bt} |\psi_D(w,\bs)| a_2(\bs,v,G)^{-1} \\
& \hspace*{50pt} \left| K^\pr \left(\frac{G_m(X_{Pi})-\bs}{h}\right) \right|
\left| \frac{G(X_{Pi})-\bs}{h}\right| |G_m(X_{Pi})-G(X_{Pi})| \, d\bs \\
\tilde{H}_b(w,v,t,X_{Pi})
& = \frac{1}{h^4} \int_0^{\bt} |\psi_D(w,\bs)| a_2(\bs,v,G)^{-1} \\
& \hspace*{50pt} |K^{\pr \pr}(\upsilon_i)|
\left|\frac{G(X_{Pi})-\bs}{h}\right| (G_m(X_{Pi})-G(X_{Pi}))^2 \\
& \hspace*{50pt} I\left( \min \left\{ \left|\frac{G(X_{Pi})-\bs}{h}\right|,
\left|\frac{G_m(X_{Pi})-\bs}{h}\right| \right\} \leq 1 \right) \, d\bs
\end{align*}
where $\upsilon_i$ is a value between $(G(X_{Pi})-\bs)/h$ and $(G_m(X_{Pi})-\bs)/h$.
Now,
$$
\tilde{H}_a(w,v,\bt,X_{Pi})
\leq O_p(1) h^{-1} \| G_m - G \|_\infty \tilde{\fA}(G(X_{Pi}))
$$
with
$$
\tilde{\fA}(\xi) = \frac{1}{h^2} \int_0^t K^\pr \left(\frac{\xi-s}{h}\right) \left|\frac{\xi-s}{h}\right| ds
$$
By the same argument as used before for $\fA(X_{Pi})$, we have
$\tilde{\fA}(G(X_{Pi})) = O_p(1)$. Thus,
$$
\tilde{H}_a(w,v,\bt,X_{Pi})
\leq O_p(1) h^{-1} \| G_m - G \|_\infty = O_p(1)
$$
By a similar argument,
$$
\tilde{H}_b(w,v,\bt,X_{Pi})
\leq O_p(1) h^{-2} \| G_m - G \|_\infty^2 = O_p(1)
$$
Hence, by the same argument as used before for  $\cB_1^*(t,u,u)$, we find that $\sqrt{mh} \, \, \Omega^*(\bt, u) = o_p(1)$.
Finally, by the same argument as used for $\cB_1^{**}(t,u)$, we obtain
$\sqrt{mh} \, \, \Omega^{**}(\bt, u) = o_p(1)$.

Finally, we have
\begin{align*}
\hat{\Lambda}(t) - \Lambda(t)
& = \hat{\Lambda}_D(G_m(t)) -  \Lambda_D(G(t)) \\
& = (\hat{\Lambda}_D(G(t)) -  \Lambda_D(G(t)))
+ (\hat{\Lambda}_D(G_m(t)) -  \hat{\Lambda}_D(G(t)))
\end{align*}
We have just shown that
$\sqrt{mh} \, \, (\hat{\Lambda}_D(G(t)) -  \Lambda_D(G(t)))$ converges in distribution to a mean-zero normal
distribution. We now show that $\sqrt{mh} \, \, (\hat{\Lambda}_D(G_m(t)) -  \hat{\Lambda}_D(G(t))) = o_p(1)$.
We have
$$
| \hat{\Lambda}_D(G_m(t)) -  \hat{\Lambda}_D(G(t)) |
\leq \| G_m - G \|_\infty \sup_{\bs \in [0,1]}
\left| \int_0^\tau \hat{\psi}_{D}(u,\bs) \hat{\Lambda}_{0,D}^*(u|\bs)  \, du \right|
$$
we know that $\| G_m - G \|_\infty = O_p(m^{-1/2})$, and we have
$$
\sup_{\bs \in [0,1]}
\left| \int_0^\tau \hat{\psi}_{D}(u,\bs) \hat{\Lambda}_{0,D}^*(u|\bs)  \, du \right| = O_p(1)
$$
so we get $\sqrt{mh} \, \,(\hat{\Lambda}_D(G_m(t)) -  \hat{\Lambda}_D(G(t))) = o_p(1)$ as desired.

\vspace*{0.5in}

\noindent
\bf \underline{Additional References} \rm

Andersen, P. K., and Gill, R. D. (1982). Cox's regression model for
counting processes: a large sample study, \textit{Annals of Statistics}
\textbf{10}: 1100--1120.

Dudley, R. M. (1999). \textit{Uniform Central Limit Theorems}. Cambridge University Press, Cambridge.

Gin\'{e}, E. and Guillou, A. (2002). Rates of strong consistency for multivariate kernel
density estimators. \textit{Annales de l'Institut Henri Poincar\'{e} (B) Probability and Statistics} \textbf{38}: 907-–922.

Schuster, E. F. (1969). Estimation of a probability density function and its derivatives.
\textit{Annals of Mathematical Statistics} \textbf{40}: 1187--1195.

van der Vaart, A. W. (1998). \textit{Asymptotic Statistics}. Cambridge University Press, Cambridge.


\begin{thebibliography}{6}
\expandafter\ifx\csname natexlab\endcsname\relax\def\natexlab#1{#1}\fi

\bibitem[{Chatterjee et~al.(2006)Chatterjee, Kalaylioglu, Shih \&
  Gail}]{Chatt2006}
\textsc{Chatterjee, N.}, \textsc{Kalaylioglu, Z.}, \textsc{Shih, J.~H.} \&
  \textsc{Gail, M.} (2006).
\newblock Case–control and case-only designs with genotype and family history
  data: estimating relative risk, residual familial aggregation, and cumulative
  risk.
\newblock \textit{Biometrics} \textbf{62}, 36--48.

\bibitem[{Dabrowska(1987)}]{Dabr1987}
\textsc{Dabrowska, D.~M.} (1987).
\newblock Non-parametric regression with censored survival time data.
\newblock \textit{Scandinavian Journal of Statistics} \textbf{14}, 181--197.

\bibitem[{Doksum et~al.(2017)Doksum, Jiang, Sun \& Wang}]{Doksum2017}
\textsc{Doksum, K.~A.}, \textsc{Jiang, J.}, \textsc{Sun, B.} \& \textsc{Wang,
  S.} (2017).
\newblock Nearest neighbor estimates of regression.
\newblock \textit{Computational Statistics and Data Analysis} \textbf{110},
  64--74.

\bibitem[{Gorfine et~al.(2017)Gorfine, Bordo \& Hsu}]{Gorfine2017}
\textsc{Gorfine, M.}, \textsc{Bordo, N.} \& \textsc{Hsu, L.} (2017).
\newblock A fully nonparametric estimator of the marginal survival function
  based on case–control clustered age-at-onset data.
\newblock \textit{Biostatistics} \textbf{18}, 76--90.

\bibitem[{Shih \& Chatterjee(2002)}]{Shih2002}
\textsc{Shih, J.~H.} \& \textsc{Chatterjee, N.} (2002).
\newblock Analysis of survival data from case–control family studies.
\newblock \textit{Biometrics} \textbf{58}, 502--509.

\bibitem[{Stanford et~al.(1999)Stanford, Wicklund, McKnight, Daling \&
  Brawer}]{Stan1999}
\textsc{Stanford, J.~L.}, \textsc{Wicklund, K.~G.}, \textsc{McKnight, B.},
  \textsc{Daling, J.~R.} \& \textsc{Brawer, M.~K.} (1999).
\newblock Vasectomy and risk of prostate cancer.
\newblock \textit{Cancer Epidemiology Biomarkers and Prevention} \textbf{8},
  881--886.

\end{thebibliography}
\end{document}